\documentclass[twocolumn]{jpsj3}
\usepackage{txfonts}
\usepackage{amsmath,amssymb}
\usepackage{bm}
\usepackage{graphicx, color}
\graphicspath{{./figures/}}
\usepackage{color}

\title{Quantum Phase Transition in Fully-Connected Quantum Wajnflasz--Pick Model}
\author{Yuya~Seki$^{1}$, Shu~Tanaka$^{2,3}$, Shiro~Kawabata$^{1}$}
\inst{$^1$Nanoelectronics Research Institute, National Institute of Advanced Industrial Science and Technology (AIST),
1-1-1 Umezono, Tsukuba, Ibaraki 305-8568 Japan\\
$^2$Green Computing Systems Research Organization, Waseda University
27, Waseda-cho, Shinjuku-ku, Tokyo, 162-0042 Japan\\
$^3$ JST PRESTO, 4-1-8 Honcho, Kawaguchi-shi, Saitama 332-0012, Japan} 
\abst{%
  We construct a quantum Wajnflasz--Pick model that is a generalized quantum Ising model,
  and investigate a nature of quantum phase transitions of the model with infinite-range interactions.
  Quantum phase transition phenomena have drawn attention in the field
  of quantum computing as well as condensed matter physics, since the phenomena are closely
  related to the performance of quantum annealing (QA) and adiabatic quantum computation (AQC).
  We add a quantum driver Hamiltonian to the Hamiltonian of classical Wajnflasz--Pick model.
  The classical Wajnflasz--Pick model consists of two-level systems as with the usual Ising model.
  Unlike the usual Ising spin, each of the upper and the lower levels of the system can be degenerate.
  The states in the upper level and the lower level are referred to as upper states and lower states, respectively.
  The quantum driver Hamiltonian we introduced causes 
  \textit{spin flip} between the upper and the lower states and \textit{state transitions}
  within each of the upper and the lower states.
  Numerical analysis showed that the model undergoes first-order phase transitions
  whereas a corresponding quantum Ising model, quantum Curie--Weiss model,
  does not undergo first-order phase transitions.
  In particular, we observed an anomalous phenomenon that the system undergoes
  successive first-order phase transitions under certain conditions.
  The obtained results indicate that the performance of QA and AQC
  by using degenerate two-level systems can be controlled by the parameters in the systems.
}
\begin{document}
\pagestyle{plain}
\maketitle
\thispagestyle{plain}

\section{Introduction}
A primary purpose of quantum annealing (QA)~\cite{kadowaki1998quantumannealing}
and adiabatic quantum computation (AQC)~\cite{farhi2000quantum,farhi2001aquantum}
are to find a ground state of a classical problem Hamiltonian that encodes a desired solution
of a combinatorial optimization problem.
Theoretical and numerical studies have revealed the advantage of QA and AQC for certain problems
over simulated annealing that is a classical counterpart
of QA and AQC~\cite{morita2008mathematical,santoro2006optimization,katzgraber2015seeking,denchev2016what,das2008colloquim,tanaka2017quantum}.
In addition to the theoretical and numerical studies, experimental effort has been devoted
to the development of quantum devices
for QA and AQC~\cite{johnson2011quantum,maezawa2017design,chen2011experimental}.
Most of the devices adopt a superconducting circuit as a unit of quantum information called qubit
that obeys quantum mechanics.
Although integration of a lot of qubits is difficult for such quantum devices,
D-Wave Systems has released a programmable device consisting of 2048 qubits,
extending range of applications of QA and AQC on actual devices.
Thus, QA and AQC are attractive computational models for state-of-the-art quantum technology.

Both of QA and AQC find a ground state of a classical Hamiltonian based on quantum dynamics under a Hamiltonian.
We consider the following standard form of the Hamiltonian for QA and AQC:
\begin{align}
  \hat{H}(s) = s \hat{H}_{\text{P}} + (1-s)\hat{V},
  \label{WP_eq:H_QA}
\end{align}
where $\hat{H}_{\text{P}}$ is a quantum Hamiltonian obtained by replacing classical variables
in the classical problem Hamiltonian with quantum operators,
and $\hat{V}$ is a quantum driver Hamiltonian that has a unique and easy-to-prepare ground state.
In addition, the quantum driver Hamiltonian $\hat{V}$ must not commute with $\hat{H}_{\text{P}}$.
The non-commutative property induces quantum fluctuations into the system, causing evolution of quantum states.
The variable $s \in [0, 1]$ is a controllable variable during a procedure of QA and AQC.
We assume that $s$ increases monotonically.
The ground state of $\hat{V}$ evolves adiabatically into the ground state of $\hat{H}_{\text{P}}$
under the Hamiltonian~\eqref{WP_eq:H_QA} if $s$ changes sufficiently slowly. 
In the present paper, we focus on AQC, which is based
on the quantum adiabatic evolution.
The adiabatic theorem ensures that the running time of QAC to get the ground state of the classical
problem Hamiltonian with a high probability scales as $1/\mathrm{Poly}(\Delta_{\text{min}})$~\cite{farhi2000quantum, morita2007convergence, jansen2007bounds}.
Here, $\mathrm{Poly}$ denotes a polynomial function, and $\Delta_{\text{min}}$ the minimum energy gap
between the instantaneous ground state and the first-excited state of the total Hamiltonian~\eqref{WP_eq:H_QA}
during the time evolution.

Quantum phase transition~\cite{sachdev2011quantum} is a crucial phenomenon in AQC as well as condensed matter physics.
A number of studies for the performance of AQC have revealed that
the minimum energy gap $\Delta_{\text{min}}$ decays exponentially for systems undergoing
a first-order phase transition, indicating the failure of AQC
that the required running time increases exponentially~\cite{znidaric2006exponential,jorg2010energy,jorg2010firstorder}.
Still, AQC is not hampered by second-order phase transitions,
because $\Delta_{\text{min}}$ decays polynomially~\cite{dusuel2005continuous, seki2012quantum}.
Although phase transitions are phenomena in the thermodynamic limit,
we can evaluate the performance of AQC on a finite size system from the phenomena.
Thus, avoiding first-order phase transitions is of great importance for AQC.
A method to avoid the problematic first-order phase transitions is constructing $\hat{V}$
so that the total system does not undergo the first-order phase transitions.
Previous studies have reported that ingenious choices of $\hat{V}$ enable us to avoid the first-order phase transitions
that is inevitable when only the conventional transverse-field term is used as the quantum driver Hamiltonian
for certain models~\cite{seki2012quantum, seoane2012many, seki2015quantum, susa2018exponential}.

The main purpose of the present paper is to
reveal effect on phase transitions in the procedure of AQC by changing the unit of quantum information.
To this end, we consider an extension of the classical Wajnflasz--Pick model~\cite{wajnflasz1971transitions}
to a quantum system.
The classical model is regarded as a generalization of Ising model.
Like an Ising spin, each variable has two distinct energy levels, namely upper and lower levels,
under a longitudinal magnetic field.
The difference from the usual Ising model is that the upper and the lower levels
are allowed to be \textit{degenerate}.
The degrees of degeneracy of the two levels differ in general.
States in the upper and the lower levels are referred to as upper states and lower states, respectively.
The classical Wajnflasz--Pick model has been studied because of its rich statistical-mechanical nature.
The model describes spin-crossover phase transitions~\cite{wajnflasz1971transitions,bousseksou1992ising,bousseksou1993ising,miyashita2010phase}
and charge transfer phase transitions~\cite{miyashita2003generalized,miyashita2005structures,tokoro2006huge,miyashita2010phase}.
An important point is that we can artificially control the order of phase transitions by adjusting parameters of the model.
The order of phase transition changes depending on the ratio of the degrees of degeneracy
owing to entropic effects.
Although the entropic effects disappear in the low-temperature limit,
we observed similar effects due to a quantum driver Hamiltonian in our model.
For the simplicity of analysis, we consider a classical fully-connected Wajnflasz--Pick model.
We constructed a fully-connected quantum Wajnflasz--Pick model by adding a non-commutative term to the Hamiltonian of classical Wajnflasz--Pick model as a quantum driver Hamiltonian.
The quantum driver Hamiltonian causes spin flip, i.e.\ transitions between the upper and the lower states,
as with the transverse-field term in the usual quantum Ising model.
In addition, the quantum driver Hamiltonian causes state transitions within the upper (lower) states.
We introduced a transition factor $\omega$ for the latter transitions as described later.

We investigated the statistical-mechanical properties of the fully-connected quantum Wajnflasz--Pick model
by using the mean-field analysis~\cite{nishimori2010elements}, since the spins are fully connected.
We derived a pseudo-free energy of the system by using the Suzuki--Trotter decomposition~\cite{suzuki1976relationship}.
The global minimum of the pseudo-free energy is the value of free energy,
and the point of the global minimum is magnetization of a stable state.
Finding all the minima of the pseudo-free energy, we investigated phase transitions in the model.
We determined the order of phase transitions from the magnetization of the stable state as a function
of the variable $s$ in the Hamiltonian~\eqref{WP_eq:H_QA}.

The main proposition of the present paper is that the order of quantum phase transitions
in the fully-connected quantum Wajnflasz--Pick model is controllable by adjusting the degrees of degeneracy of
the upper and the lower levels, and the transition factor $\omega$.
We derived a Hamiltonian of quantum Ising model whose statistical-mechanical properties are equivalent
to the fully-connected quantum Wajnflasz--Pick model for $\omega \ge 0$.
From the results, we obtained conditions that the system undergoes a first-order phase transition
when $\omega \ge 0$.
The first-order phase transition occurs when an effective longitudinal field of the quantum Ising model
changes its sign in a region with weak quantum fluctuations.
Our numerical analysis showed the existence of first-order phase transitions for negative $\omega$ too.
A noteworthy result is that the fully-connected quantum Wajnflasz--Pick model
undergoes a first-order quantum phase transition twice for certain parameters.
The anomalous phenomena are not seen in the corresponding quantum Ising model.

The present paper is organized as follows.
In Sec.~\ref{sec:model}, we introduce the classical Wajnflasz--Pick model, and explain
how to extend the model to a quantum model.
In the section, we investigate the ground state of a initial Hamiltonian of the quantum model
in order to reveal a difference from the transverse-field term in the usual quantum Ising model.
Section~\ref{sec:analysis} is devoted to explain methods to derive the pseudo-free energy and find all the minima of the pseudo-free energy.
In Sec.~\ref{WP_sec:results}, we first describe the results of the number of first-order phase transitions
for $g_{\text{u}} > g_{\text{l}}$ in Sec.~\ref{WP_sec:sub_upper-upper} to Sec.~\ref{WP_sec:sub_lower-lower}.
We demonstrate the equivalent Hamiltonian of quantum Ising model here.
We next describe results for the balanced case $g_{\text{u}} = g_{\text{l}}$ in Sec.~\ref{WP_sec:sub_results_for_balanced}.
We conclude in Sec.~\ref{sec:conclusion}.

\section{Model}
\label{sec:model}

\subsection{Classical Wajnflasz--Pick Model}
\label{sec:class-wajnfl-pick}

\begin{figure}[tp]
  \centering
  \includegraphics[width=60mm]{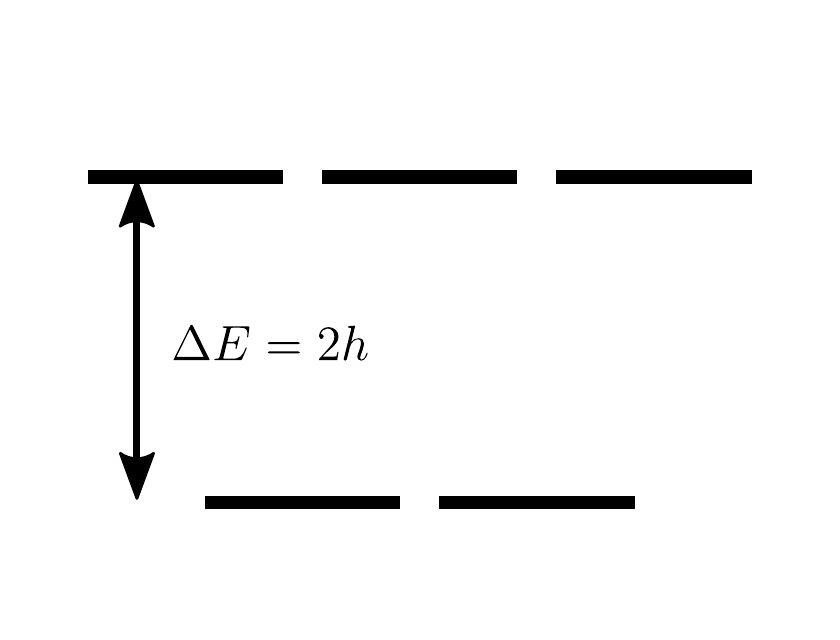}
  \caption{Schematic picture of the energy level of the variable $\tau^{z}$ with $g_{\text{u}} = 3$ and $g_{\text{l}} = 2$
  subjected to a positive longitudinal field $h$.
  The upper three horizontal lines represent the upper states,
  and the lower two horizontal lines the lower states.
  }
  \label{WP_fig:spin_classical_WP}
\end{figure}

Before moving onto the explanation for the fully-connected quantum Wajnflasz--Pick model
that we discuss in the present paper, let us introduce the classical counterpart
in order to clarify the difference from the usual Ising models.
The classical Wajnflasz--Pick model is considered as a generalization of the Ising model:
The model is composed of two-level variables like an Ising spin,
but each level can be degenerate.
Figure~\ref{WP_fig:spin_classical_WP} shows the energy levels of the variable
with the three-fold degenerate upper level and the two-fold degenerate lower level as an example.
In the present paper, we consider the fully-connected Wajnflasz--Pick model for the simplicity of analysis.
The Hamiltonian is defined as
\begin{align}
  H &= -\frac{1}{N}\sum_{i,j=1}^{N}\tau_{i}\tau_{j} - h \sum_{i=1}^{N} \tau_{i},
      \label{WP_eq:Classical_H}
\end{align}
where $N$ represents the number of sites and $h$ a longitudinal field.
The variable $\tau_{i}$ takes $\pm 1$ for all $i$.
The state of $\tau_{i} = +1$ is referred to as an upper state, and the state $\tau_{i} = -1$
as a lower state.
The degree of degeneracy of the upper states is denoted by $g_{\text{u}}$,
and that of lower sates is denoted by $g_{\text{l}}$.

\begin{figure}[tp]
  \centering
  \includegraphics{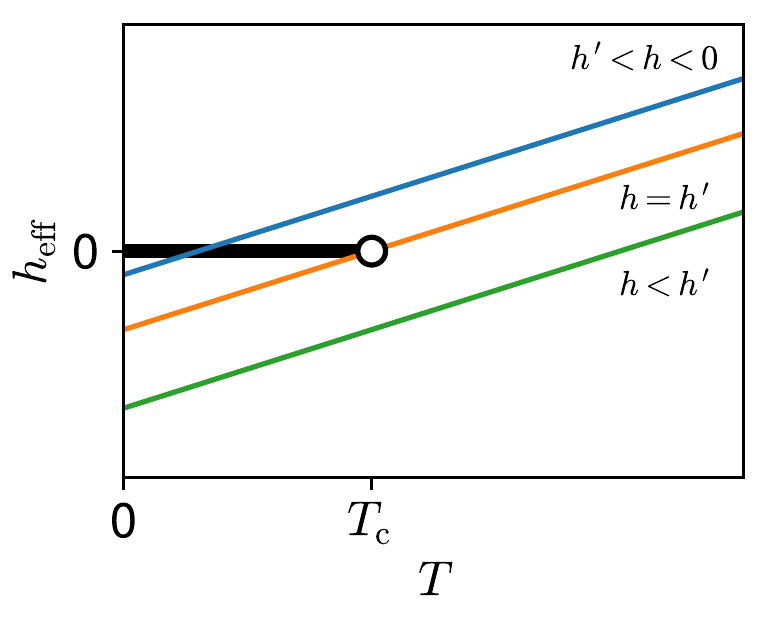}
  \caption{(Color online). Phase diagram of the Ising model~\eqref{WP_eq:Classical_H_Ising}
  on a $(T, h_{\text{eff}})$ plane,
  and annealing paths for $g_{\text{u}} = 2$ and $g_{\text{u}} = 1$.
  The white circle represents the critical point $T_{\text{c}}$ and $h_{\text{eff}, \text{c}} = 0$, where
  $T_{\text{c}}$ denotes the Curie temperature.
  The thick solid black line represents the first-order phase boundary.
  The annealing represented as the thin solid orange line crosses the critical point
  when $h = h' = -(T_{\text{c}}/2) \ln (g_{\text{u}}/g_{\text{l}})$, meaning that
  the system undergoes a second-order phase transition.
  For $h' < h < 0$, a first-order phase transition occurs, since the annealing path (thin solid blue line)
  goes across the first-order phase boundary.
  No phase transitions occur when $h < h'$ as shown by the thin solid green line.
  }
  \label{WP_fig:path_on_pd_classical_WP}
\end{figure}

Let us consider phase transitions of the model~\eqref{WP_eq:Classical_H}
for fixed $g_{\text{u}}$, $g_{\text{l}}$, and $h$.
The model can be reduced to the following Ising spin system~\cite{miyashita2005structures}:
\begin{align}
  H' &= -\frac{1}{N}\sum_{i,j=1}^{N} \sigma_{i}\sigma_{j}
       - \biggl( h + \frac{T}{2}\ln \frac{g_{\text{u}}}{g_{\text{l}}} \biggr)\sum_{i=1}^{N}\sigma_{i},
       \label{WP_eq:Classical_H_Ising}
\end{align}
where $\sigma_{i}$ denotes an Ising spin on site $i$.
This Hamiltonian yields the same partition function
as that of the system~\eqref{WP_eq:Classical_H} at a temperature $T$.
Hence, the phase diagram is obtained from the knowledge of the usual Ising model.
The difference from the usual Ising model is
that the effective longitudinal field
\begin{align}
  h_{\text{eff}} &= h + \frac{T}{2}\ln \frac{g_{\text{u}}}{g_{\text{l}}}
  \label{WP_eq:effective-h}
\end{align}
depends on a temperature.
The second term in the right hand side of Eq.~\eqref{WP_eq:effective-h} is a bias caused by entropic effects,
since more degenerate states are preferred due to the term.
Figure~\ref{WP_fig:path_on_pd_classical_WP} shows a phase diagram of the Ising model~\eqref{WP_eq:Classical_H_Ising}
on the $(T, h_{\text{eff}})$ plane.
A critical point of the system~\eqref{WP_eq:Classical_H_Ising} is given as
$T=T_{\text{c}}$ and $h_{\text{eff},\text{c}} = 0$, where $T_{\text{c}}$ denotes the critical temperature
of the Curie--Weiss model.
We refer to trajectories of the parameters during an annealing procedure as \textit{annealing paths}
in this section.
Some annealing paths are plotted in Fig.~\ref{WP_fig:path_on_pd_classical_WP} as thin lines.
The annealing path for $h = h' = -(T_{\text{c}}/2) \ln (g_{\text{u}}/g_{\text{l}})$ crosses the critical point.
Hence, the system undergoes a second-order phase transition during the annealing procedure represented by the annealing path.
The system does not undergo any phase transitions for $h$ smaller than $h'$.
When $h' < h < 0$, the annealing path crosses the first-order phase boundary,
indicating that the system undergoes a single first-order phase transition.
Thus, the order of phase transition of the classical Wajnflasz--Pick model is controllable
by adjusting the ratio of the degrees of degeneracies and a longitudinal field.

In the present paper, we are interested in quantum phase transitions in the low-temperature limit.
Although the entropic effect disappears in the low-temperature limit,
we observed similar effects due to a quantum driver Hamiltonian that we introduce in the next section.

\subsection{Fully-Connected Quantum Wajnflasz--Pick Model}

\begin{figure}[tp]
  \centering
  \includegraphics[width=60mm]{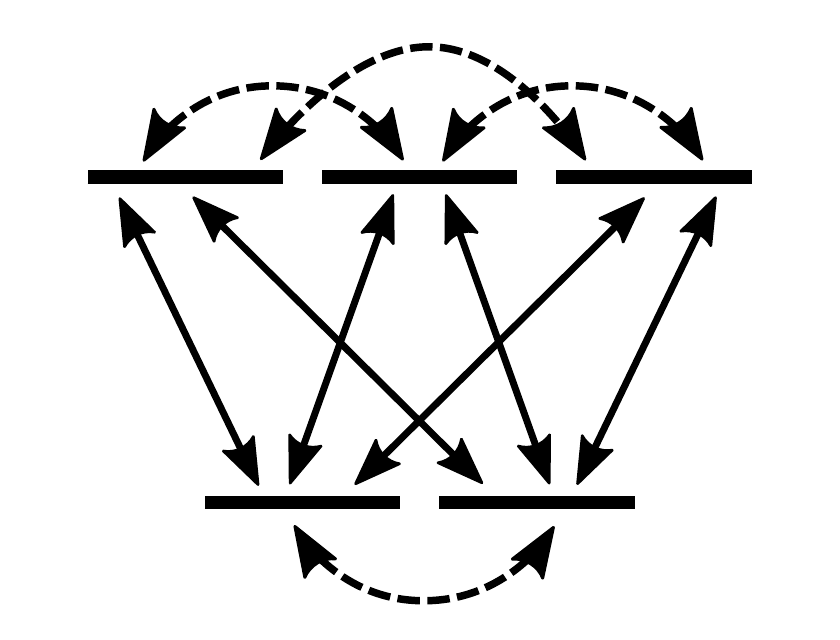}
  \caption{Schematic picture of the upper states and the lower states of a single variable
  with $g_{\text{u}} = 3$ and $g_{\text{l}} = 2$,
  and state transitions caused by the operator $\tau^{x}$ in the fully-connected quantum Wajnflasz--Pick model.
  The upper horizontal lines and lower horizontal lines represent
  the upper states and the lower states, respectively.
  The solid arrows represents state transitions called spin flip resulting from
  the off-diagonal block, $\mathbf{1}$, of the operator $\tau^{x}$.
  The dashed arrows represents state transitions within each of the upper and the lower states.
  The transitions within the states are caused by the diagonal block, $\boldsymbol{\Omega}$,
  of the operator $\tau^{x}$.
  Note that the positions of the states in the vertical direction no longer indicate
  the energy levels because of the transitions.
  }
  \label{WP_fig:spin_quantum_WP}
\end{figure}

The Hamiltonian of the fully-connected quantum Wajnflasz--Pick model is defined as
\begin{align}
  \hat{H}(s) &= s\left[-\frac{1}{N} \sum_{i,j=1}^{N} \hat{\tau}_{i}^{z}\hat{\tau}_{j}^{z}
               - h \sum_{i=1}^{N}\hat{\tau}_{i}^{z}\right]
               - (1-s) \sum_{i=1}^{N}\hat{\tau}_{i}^{x},
               \label{WP_eq:quantum_WP}
\end{align}
where $s \in [0, 1]$ denotes the controllable variable increasing monotonically with a procedure of AQC,
and $h$ a longitudinal field,
which is a constant during the procedure.
The operator $\hat{\tau}^{z}$ represents a two-level system with degeneracy:
\begin{align}
  \hat{\tau}^{z} &\equiv \mathrm{diag} (\underbrace{+1, \dotsc, +1}_{g_\text{u}},
                   \underbrace{-1, \dotsc, -1}_{g_\text{l}}).
\end{align}
Here, $g_{\text{u}}$ and $g_{\text{l}}$ represent a degree of degeneracy
of the upper states and that of the lower states, respectively.
We fix both the parameters during the procedure.
The last term in Eq.~\eqref{WP_eq:quantum_WP} is a driver part
that induces quantum fluctuations into the system.
In the present paper, we define the operator $\hat{\tau}^{x}$ as follows:
\begin{align}
  \hat{\tau}^{x} &\equiv \frac{1}{c}
                   \begin{pmatrix}
                     \boldsymbol{\Omega}(g_{\text{u}})               & \mathbf{1}(g_{\text{u}}, g_{\text{l}}) \\
                     \mathbf{1}(g_{\text{l}}, g_{\text{u}}) & \boldsymbol{\Omega}(g_{\text{l}})
                   \end{pmatrix},
                   \label{WP_eq:def_taux}
\end{align}
where the matrix $\boldsymbol{\Omega}(l)$ is a square matrix of order $l$ whose diagonal elements
are zero, and off-diagonal elements are $\omega$:
\begin{align}
  \boldsymbol{\Omega}(l) &\equiv
              \begin{pmatrix}
                0                 & \omega & \cdots            & \omega \\
                \omega & 0      & \ddots            & \vdots \\
                \vdots            & \ddots & \ddots            & \omega \\
                \omega & \cdots & \omega & 0
              \end{pmatrix}.
\end{align}
Here, $\omega$ is a real variable.
The other matrix $\mathbf{1}(n,m)$ is an $n \times m$ matrix with ones.
The off-diagonal block matrix $\mathbf{1}$ represents spin flip, and $\boldsymbol{\Omega}$
the transition within upper (lower) states.
The transitions for the case with $g_{\text{u}} = 3$ and $g_{\text{l}} = 2$
are shown in Fig.~\ref{WP_fig:spin_quantum_WP}.
The factor $c$ is a normalization factor so that the spectral norm of the operator $\hat{\tau}^{x}$
coincides with that of the operator $\hat{\tau}^{z}$.
Without loss of generality, we assume $g_{\text{u}} \ge g_{\text{l}}$.

\subsection{Properties of the quantum driver Hamiltonian}
\label{WP_sec:sub_properties_of_the_quantum_driver_H}

\begin{figure}[tp]
  \centering
  \includegraphics{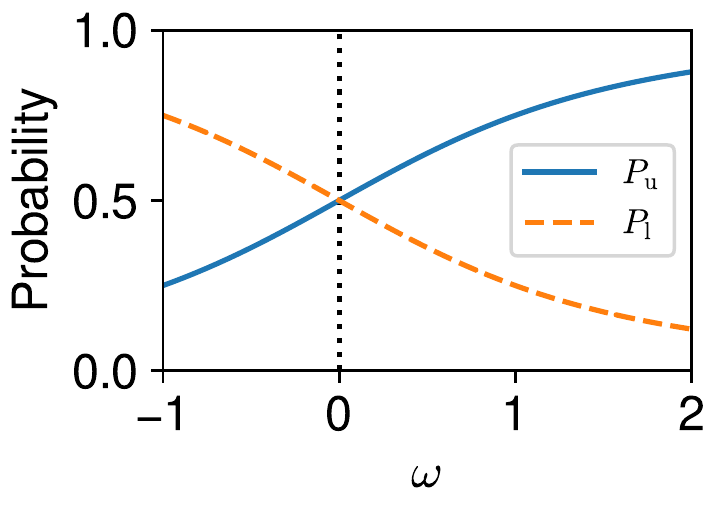}
  \caption{(Color online). The probabilities of the upper states (solid blue line)
  and the lower states (dashed orange line) in the ground state of $-\hat{\tau}^{x}$
  for $g_{\text{u}} = 3$ and $g_{\text{l}}=2$.
  The probabilities are equal when $\omega = 0$ as with the case of a transverse field of the usual quantum Ising model.
  The upper (lower) states are favored when $\omega$ is positive (negative).}
  \label{WP_fig:p_initialH}
\end{figure}

Unlike a transverse field of the usual quantum Ising model,
the probability of the upper states is not necessarily
equal to that of the lower states in the ground state of the initial Hamiltonian~\eqref{WP_eq:quantum_WP}.
The operator $-\hat{\tau}^{x}$ has $(g_{\text{u}} + g_{\text{l}} - 2)$-fold degenerate eigenvalue $\omega$.
The other eigenvalues are
\begin{align}
  \lambda_{\pm} = \left( 1 - \frac{g_{\text{u}} + g_{\text{l}}}{2}\right) \omega
                  \pm \frac{1}{2}\sqrt{(g _{\text{u}} - g _{\text{l}})^{2}\omega^{2} + 4 g_{\text{u}}g_{\text{l}}}.
\end{align}
The lowest eigenvalue is $\omega$ for $\omega < -1$, and $\lambda_{-}$ for $\omega > -1$.
Since AQC assumes a non-degenerate ground state of quantum driver Hamiltonian,
we consider only the region $\omega > -1$.
The eigenvector belonging to $\lambda_{-}$ is given as a vector consisting of two sections,
the first $g_{\text{u}}$ elements and the remaining $g_{\text{l}}$ elements:
\begin{align}
  \boldsymbol{v} &= (\underbrace{a, \dotsc , a}_{g_{\text{u}}},
  \underbrace{1, \dotsc , 1}_{g_{\text{l}}})^{\top},
\end{align}
where
\begin{align}
  a = \frac{1}{2 g_{\text{u}}}
  \left\{
    (g_{\text{u}} - g_{\text{l}} )\omega
    + \sqrt{(g_{\text{u}} - g_{\text{l}} )^{2}\omega^{2} + 4g_{\text{u}}g_{\text{l}}}
  \right\}.
\end{align}
Hence, the probability that the ground state is in the upper states is
$P_{\text{u}} = g_{\text{u}}a^{2}/(g_{\text{u}}a^{2} + g_{\text{l}})$,
and the probability for the lower states is $P_{\text{l}} = g_{\text{l}}/(g_{\text{u}}a^{2} + g_{\text{l}})$.
When the degeneracies of the upper states and lower states are the same, $g_{\text{u}} = g_{\text{l}}$,
we obtain $P_{\text{u}} = P_{\text{l}}$.
When the degeneracies are not equal, the probabilities vary depending on $\omega$.
Figure~\ref{WP_fig:p_initialH} shows the probabilities
$P_{\text{u}}$ and $P_{\text{l}}$ for $g_{\text{u}} = 3$ and $g_{\text{l}}=2$
as functions of $\omega$.
We can see that the probabilities are equal when $\omega = 0$, $P_{\text{u}} > P_{\text{l}}$ when $\omega > 0$,
and $P_{\text{u}} < P_{\text{l}}$ when $\omega < 0$.
The same properties hold for other values of $g_{\text{u}}$ and $g_{\text{l}}$ ($< g_{\text{u}}$).
We thus control the probabilities in the ground state of the initial Hamiltonian~\eqref{WP_eq:quantum_WP}
by adjusting $\omega$.

\section{Analysis method}
\label{sec:analysis}
We derived the pseudo-free energy to investigate an order parameter of the system.
Since the system is fully connected model, the mean-field analysis offers an effective way to study statistical-mechanical properties of the system.
We followed a canonical way to analyze quantum mean-field models
(see Appendix\ref{appdx:pseudo-free-energy} for the detailed calculation).
First, we transformed the partition function of the quantum system to a corresponding classical system
by using the Suzuki--Trotter decomposition~\cite{suzuki1976relationship}.
Next, we used the mean-field analysis, and introduced an order parameter similar to magnetization.
Applying the static ansatz that removes imaginary-time dependence of the order parameter,
we obtained the pseudo-free energy:
\begin{align}
  f &= s m^{2} - \frac{1}{\beta} \ln \Tr \exp \beta (\tilde{m} \hat{\tau}^{z} + (1-s) \hat{\tau}^{x}),
  \label{WP_eq:pf}
\end{align}
where $\beta$ represents the inverse temperature,
$m$ the magnetization that is the fraction of variables in the upper states,
and $\tilde{m}$ is defined as $\tilde{m} = s(2m+h)$.
Since we focus on the quantum phase transitions in the model, we took the low-temperature limit.
We then obtained
\begin{align}
  f &= s m^{2} - \lambda_{\text{max}},
  \label{WP_eq:pf_low_temp}
\end{align}
where $\lambda_{\text{max}}$ denotes the largest eigenvalue of the operator $\tilde{m} \hat{\tau}^{z} + (1-s) \hat{\tau}^{x}$.

We numerically calculated the all minima of the pseudo-free energy~\eqref{WP_eq:pf_low_temp}
as a function of $m$ for fixed $s$ and $h$.
We used Python packages in order for the numerical analysis
and plot of figures~\cite{oliphant2006guide,scipy,hunter2007}.
The maximum eigenvalue was calculated through the Pal-Walker-Kahan variant of the QR algorithm.
Minima of the pseudo-free energy~\eqref{WP_eq:pf_low_temp} were searched
by a brute-force method followed by the Brent's method to improve the accuracy.
The brute-force search were performed for evenly spaced values of $m$ from $-1.02$ to $1.02$, and the step size is 0.005.
The relative error of each minimum point is $1.48\times 10^{-8}$ after applying the Brent's method.
The global minimum point of the pseudo-free energy is the magnetization of a stable state for the fixed variables $s$ and $h$,
and the other minima of the pseudo-free energy, if any, are magnetization associated with metastable states.
We detected first-order phase transitions as follows.
We calculated the minima of the pseudo-free energy for various values of $s$
with a fixed value of $h$.
In the present paper, the smallest value of $s$ is $0.001$, and the largest is unity.
We considered that the system undergoes a first-order phase transition at a point $s_{\ast}$
when the difference between adjacent magnetization of the stable state at $s_{\ast}$
exceeds a threshold $0.3$, and a metastable state exists at $s_{\ast}$.
Although our method cannot detect first-order phase transitions
when the difference of adjacent magnetization is smaller than the threshold,
out method revealed qualitative behavior of the fully-connected quantum Wajnflasz--Pick model as will be shown in Sec.~\ref{WP_sec:results}.

In addition to the numerical study, we analytically investigated the fully-connected quantum Wajnflasz--Pick model
for $\omega \ge 0$.
We can get a closed form of the pseudo-free energy~\eqref{WP_eq:pf_low_temp}.
We derived a Hamiltonian of quantum Ising model whose pseudo-free energy is the same as Eq.~\eqref{WP_eq:pf_low_temp}.
The phase transition phenomena of the fully-connected quantum Wajnflasz--Pick model
can be investigated from the knowledge of the phase diagram of the quantum Ising model.

\section{Results}
\label{WP_sec:results}
Let us first consider the unbalanced case $g_{\text{u}} > g_{\text{l}}$, where the probabilities of the upper states
and the lower states in the ground state of the quantum driver Hamiltonian,
i.e. the initial Hamiltonian $\hat{H}(0)$, depend on $\omega$.
We classified our results into four groups according to whether the upper states are favored or not
for each of the quantum driver Hamiltonian and the classical problem Hamiltonian
(or equivalently the final Hamiltonian $\hat{H}(1)$).
As mentioned in Sec.~\ref{WP_sec:sub_properties_of_the_quantum_driver_H},
the upper states are favored when $\omega > 0$ in the ground state of the quantum driver Hamiltonian,
and the lower states are favored when $-1 < \omega < 0$.
Regarding the final Hamiltonian, all the variables in the ground state are in the upper states,
$\tau_{i}^{z} = 1$ for all $i$, when $h > 0$, and in the lower states, $\tau_{i}^{z} = -1$ for all $i$, when $h < 0$.
Next, we consider the balanced case $g_{\text{u}} = g_{\text{l}}$ in a similar manner as the balanced case.

Firstly, we show results for the case of $\omega \ge 0$ and $h \ge 0$, where the upper states are favored in both the initial and final ground states in Sec.~\ref{WP_sec:sub_upper-upper}.
We found that no first-order phase transitions occur in this case.
Secondly, we present results for $\omega \ge 0$ and $h < 0$ in Sec.~\ref{WP_sec:sub_upper-lower},
where the upper states are favored in the initial ground state,
and the lower states are favored in the final ground states.
We observed a single first-order phase transition that the magnetization jumps from a positive value to a negative value
in a quantum annealing procedure under certain parameters.
It should be noted that a closed form of the pseudo-free energy for the fully-connected quantum Wajnflasz--Pick model
can be obtained for the above two cases.
We derived a Hamiltonian of quantum Ising model that yields the same pseudo-free energy as that of the fully-connected quantum Wajnflasz--Pick model, then determined critical points analytically.
We confirmed that the analytical results coincides with numerical results.
Thirdly, we show results for $-1 < \omega < 0$ and $h\ge 0$ in Sec.~\ref{WP_sec:sub_lower-upper}.
The lower states and the upper states are favored in the ground initial and the final ground states, respectively,
We observed that the model occurs a single first-order phase transition in a quantum annealing procedure
like the previous case with $\omega \ge 0$.
Nevertheless, the cause of the first-order phase transition differs from the previous case.
Fourthly, we show results for $-1 < \omega < 0$ and $h<0$, where the lower states are favored in both the initial and final ground states in Sec.~\ref{WP_sec:sub_lower-lower}.
We observed anomalous phenomena that successive first-order phase transitions occur in a quantum annealing procedure
when $g_{\text{l}}$ is unity, and $-1 < \omega < 0$.
Finally, we show results for the balanced case $g_{\text{u}} = g_{\text{l}}$.
We show results for the balanced case in a similar manner to the balanced case.

\subsection{Case: $\omega \ge 0$ and $h \ge 0$}
\label{WP_sec:sub_upper-upper}

We derived a Hamiltonian of quantum Ising model equivalent to the fully-connected quantum Wajnflasz--Pick model for $\omega \ge 0$
in a sense that both the systems have the same statistical-mechanical properties.
We calculated analytically the maximum eigenvalue $\lambda_{\text{max}}$ in Eq.~\eqref{WP_eq:pf_low_temp}
by using the Perron--Frobenius theorem (see Appendix\ref{appdx:pseudo-free-energy} for detailed calculation):
\begin{align}
  \lambda_{\text{max}}
  &= \frac{1}{2} \left\{ (g_{\text{u}} + g_{\text{l}} - 2)\frac{(1-s)}{c}\omega \right.\notag \\
                       &+\left.
                         \sqrt{\left\{ (g_{\text{u}} - g_{\text{l}})\frac{1-s}{c}\omega + 2 \tilde{m} \right\}^{2}
                         + 4 g_{\text{u}}g_{\text{l}}\left(\frac{1-s}{c}\right)^{2}}\right\},
  \label{WP_eq:lambda_max}
\end{align}
where $c$ denotes the normalization factor in Eq.~\eqref{WP_eq:def_taux}.
Substituting Eq.~\eqref{WP_eq:lambda_max} to Eq.~\eqref{WP_eq:pf_low_temp}, we have
\begin{align}
  f = sm^{2} + \sqrt{\left\{ \tilde{m} + (g_{\text{u}} - g_{\text{l}})\frac{1-s}{2c}\omega \right\}^{2}
   + g_{\text{u}}g_{\text{l}}\left(\frac{1-s}{c}\right)^{2}},
\end{align}
Here, $\tilde{m} = s(2m+h)$ as described in Sec.~\ref{sec:analysis}.
We omitted a constant term with regards to $m$, since the term does not change magnetization
for either stable states or metastable states.
The pseudo-free energy is the same as that of quantum Ising model whose Hamiltonian is given as
\begin{align}
  \hat{H}' = s\left[ -\frac{1}{N}\sum_{i,j = 1}^{N}\hat{\sigma}_{i}^{z}\hat{\sigma}_{j}^{z}
  - h_{\text{eff}} \sum_{i=1}^{N}\hat{\sigma}_{i}^{z} \right]
  - \sqrt{g_{\text{u}}g_{\text{l}}}\frac{1-s}{c}\sum_{i=1}^{N}\hat{\sigma}_{i}^{x},
  \label{WP_eq:H_equiv_spin-1/2}
\end{align}
with an effective longitudinal field
\begin{align}
  h_{\text{eff}} = h + \frac{1-s}{s}\frac{g_{\text{u}}-g_{\text{l}}}{2c}\omega.
  \label{WP_eq:quantum_heff}
\end{align}
Here, $\hat{\sigma}_{i}^{z}$ and $\hat{\sigma}_{i}^{x}$ are $z$ and $x$ components of the Pauli matrices
acting on the site $i$, respectively.
We describe the derivation of the pseudo-free energy of Eq.~\eqref{WP_eq:H_equiv_spin-1/2}
in Appendix\ref{appdx:equivalent_spin-1/2}.
Since Eq.~\eqref{WP_eq:H_equiv_spin-1/2} is the Hamiltonian of the quantum Curie--Weiss model,
a critical point of the system is given as
\begin{align}
  s_{\text{c}} = \frac{\sqrt{g_{\text{u}}g_{\text{l}}}}{\sqrt{g_{\text{u}}g_{\text{l}}} + 2c} \quad \text{and}\quad
  h_{\text{eff}, \text{c}} = 0.
  \label{WP_eq:condition_ciritcal_point}
\end{align}
A first-order phase transition occurs when the effective longitudinal field changes its sign
within the region $s_{\text{c}} < s \le 1$.

\begin{figure}[tp]
  \centering
  \includegraphics{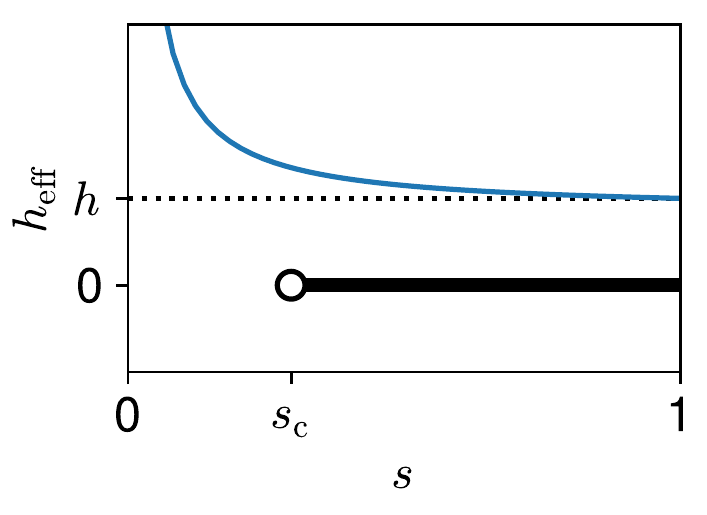}
  \caption{(Color online). Phase diagram of the quantum Ising model~\eqref{WP_eq:H_equiv_spin-1/2}
  and quantum annealing path for positive $\omega$ and $h$.
  The white circle represents the critical point $(s_{\text{c}}, h_{\text{eff}, \text{c}})$,
  and thick solid black line the first-order phase boundary.
  The thin solid blue line is a quantum annealing path.
  The quantum annealing path crosses neither the critical point nor the first-order phase boundary
  if $h \ge 0$.}
  \label{WP_fig:path_on_pd_positive_omega_positive_h}
\end{figure}

We revealed quantum phase transitions in the fully-connected quantum Wajnflasz--Pick model
from the phase diagram of the equivalent quantum Ising model.
Figure~\ref{WP_fig:path_on_pd_positive_omega_positive_h} represents a phase diagram
of the equivalent quantum Ising model on a $(s, h_{\text{eff}})$ plane.
The thin solid blue line in Fig.~\ref{WP_fig:path_on_pd_positive_omega_positive_h}
is a trajectory of the effective longitudinal field
during a quantum annealing procedure, where $s$ increases from zero to unity monotonically.
In the present paper, we refer to such a trajectory on the plane as a \textit{quantum annealing path}.
Considering that all the variables except $s$ in the right-hand side of Eq.~\eqref{WP_eq:quantum_heff} are fixed
during the quantum annealing procedure, $h_\text{eff}$ decreases monotonically,
and $h_\text{eff} = h$ ($\ge 0$) at $s = 1$.
Since the quantum annealing paths do not go through either the critical point or the first-order phase boundary,
the system~\eqref{WP_eq:H_equiv_spin-1/2} does not undergo phase transitions.
Consequently, no phase transitions occur in the fully-connected quantum Wajnflasz--Pick model for $\omega \ge 0$ and $h \ge 0$.
We confirmed the results from the numerical analysis as well.

\subsection{Case: $\omega \ge 0$ and $h<0$}
\label{WP_sec:sub_upper-lower}

\begin{figure}[tp]
  \centering
  \includegraphics{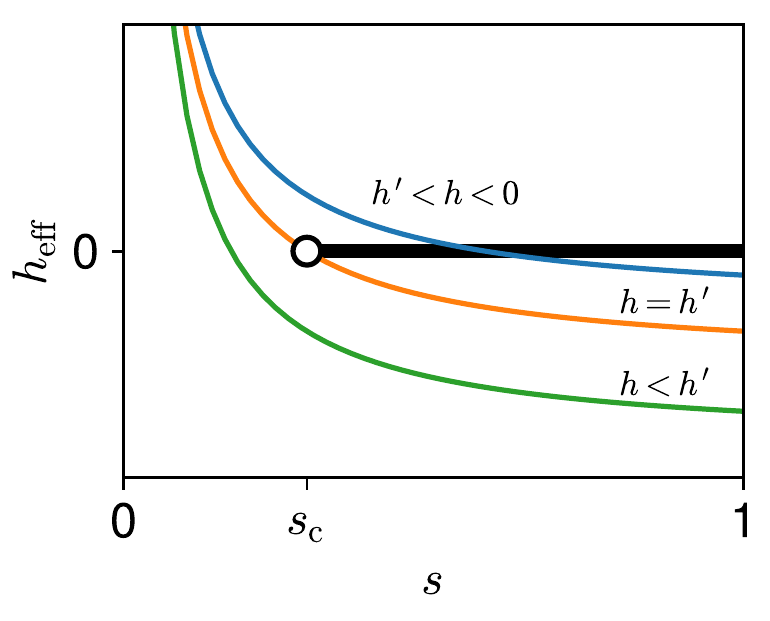}
  \caption{(Color online). Phase diagram of the quantum Ising model~\eqref{WP_eq:H_equiv_spin-1/2}
  and quantum annealing paths for positive $\omega$ and some negative $h$'s.
  The white circle represents the critical point $(s_{\text{c}}, h_{\text{eff}, \text{c}})$,
  and thick solid black line the first-order phase boundary.
  The thin solid lines are quantum annealing paths for $h' < h < 0$ (blue), $h = h'$ (orange), and $h < h'$ (green).
  Since the blue line crosses the black line, a first-order phase transition occurs in the quantum annealing procedure for the annealing path.
  The orange line passes through the critical point, indicating that the system undergoes a second-order phase transition.
  No phase transitions occur in a quantum annealing procedure represented by the green line.
  }
  \label{WP_fig:path_on_pd_positive_omega_negative_h}
\end{figure}

We can investigate phase transition phenomena for $\omega \ge 0$ and $h<0$ in the same manner
as the previous results in Sec.~\ref{WP_sec:sub_upper-upper}.
Figure~\ref{WP_fig:path_on_pd_positive_omega_negative_h} shows the phase diagram
of the quantum Ising model~\eqref{WP_eq:H_equiv_spin-1/2} and quantum annealing paths.
Since the effective longitudinal field~\eqref{WP_eq:quantum_heff} changes from a positive to a negative value,
an quantum annealing path can cross phase transition points.
From Eqs.~\eqref{WP_eq:quantum_heff} and \eqref{WP_eq:condition_ciritcal_point},
we obtain a condition that the system undergoes a second-order phase transition:
\begin{align}
  h' = -\frac{g_{\text{u}} - g_{\text{l}}}{\sqrt{g_{\text{u}}g_{\text{l}}}} \omega.
  \label{WP_eq:condition_h_critical_point}
\end{align}
Since we assumed that $g_{\text{u}} > g_{\text{l}}$, $h'$ is negative.
Figure~\ref{WP_fig:path_on_pd_positive_omega_negative_h} indicates
that the system undergoes a single first-order phase transition
when $h' < h < 0$, and no phase transitions when $h < h'$.

\begin{figure}[tp]
  \centering
  \includegraphics{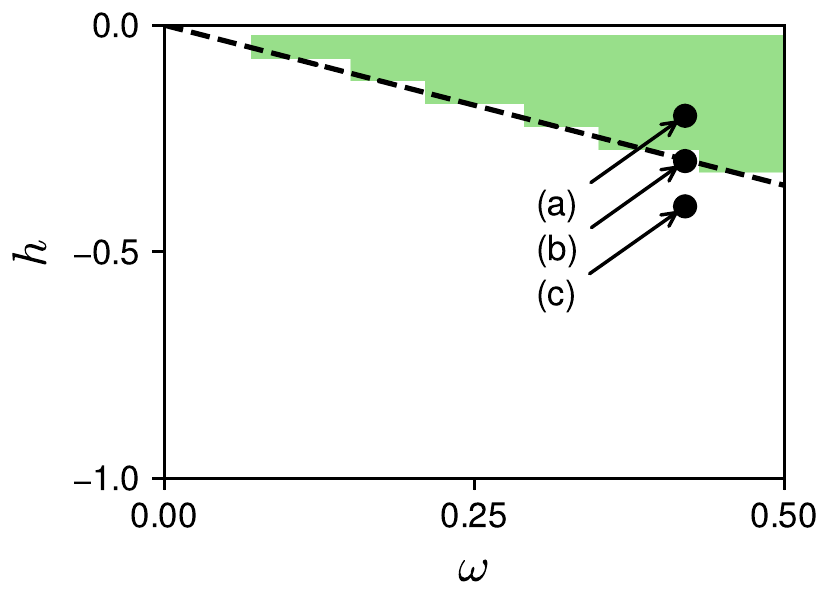}
  \caption{(Color online). The number of first-order phase transitions during quantum annealing of the fully-connected quantum Wajnflasz--Pick model~\eqref{WP_eq:quantum_WP}.
  The degrees of degeneracies are $g_{\text{u}} = 2$ and $g_{\text{l}} = 1$.
  A first-order phase transition occurs once in the light green.
  In the white region, no first-order phase transitions are observed, i.e., the system undergoes
  second-order phase transitions or no phase transitions.
  The dashed line represents the critical condition~\eqref{WP_eq:condition_h_critical_point}.
  The dashed line agrees well with the boundary between the light green region and white region,
  which is consistent with the analytical results that the system undergoes the first-order phase transition once
  when $h' < h < 0$.
  The dots are reference points where we show behavior of magnetization as a function of $s$ in Fig.~\ref{WP_fig:m_positive_omega_gu2gl1}.
  The reference points (a), (b), and (c) are $(0.42, -0.2)$, $(0.42, -0.3)$, and $(0.42, -0.4)$, respectively.
  }
  \label{WP_fig:num_pt_positive_omega_negative_h}
\end{figure}

We confirmed the above results using the numerical method.
We investigated first-order phase transitions of the model~\eqref{WP_eq:quantum_WP}
by the method described in Sec.~\ref{sec:analysis}.
Let us focus on results for $g_{\text{u}} = 2$ and $g_{\text{l}} = 1$.
Figure~\ref{WP_fig:num_pt_positive_omega_negative_h} represents the number of first-order phase transitions
occurring in a quantum annealing procedure for each fixed point of parameters $(\omega, h)$.
The critical condition~\eqref{WP_eq:condition_ciritcal_point} is denoted as the dashed line
in Fig.~\ref{WP_fig:num_pt_positive_omega_negative_h}.
The region where a single first-order phase transition occurs filled in light green
is coincide with the analytical results that the first-order phase transition occurs when $h' < h < 0$.
Although we showed results only for $g_{\text{u}} = 2$ and $g_{\text{l}} = 1$,
we observed similar results for other $g_{\text{u}}$ and $g_{\text{l}}$.

\begin{figure}[tp]
  \centering
  \includegraphics[width=70mm]{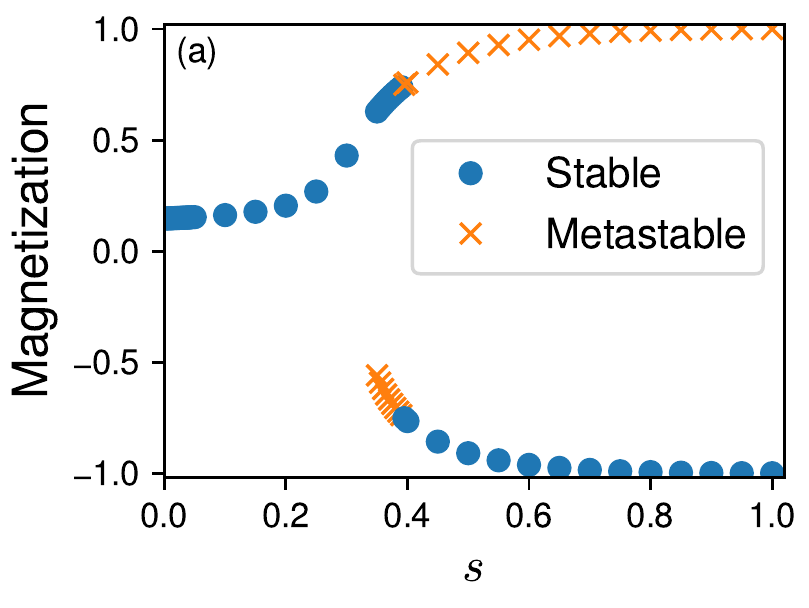}
  \includegraphics[width=70mm]{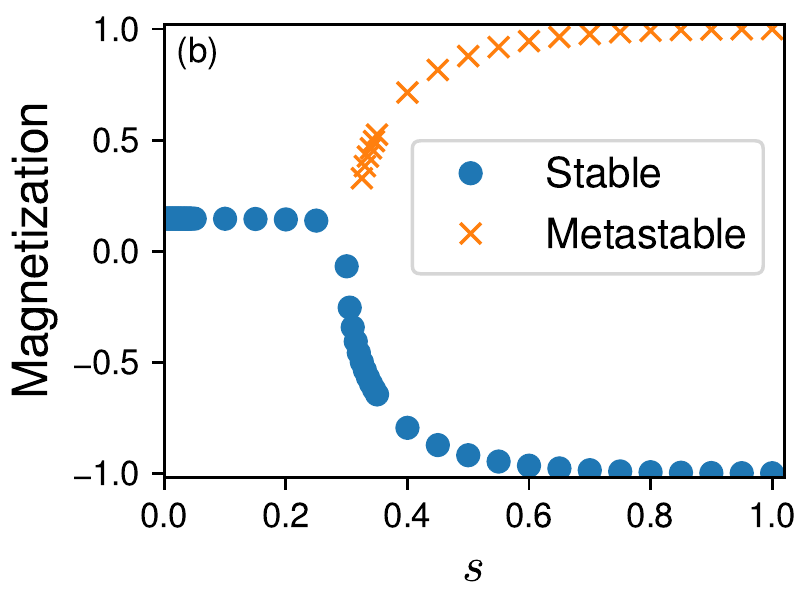}
  \includegraphics[width=70mm]{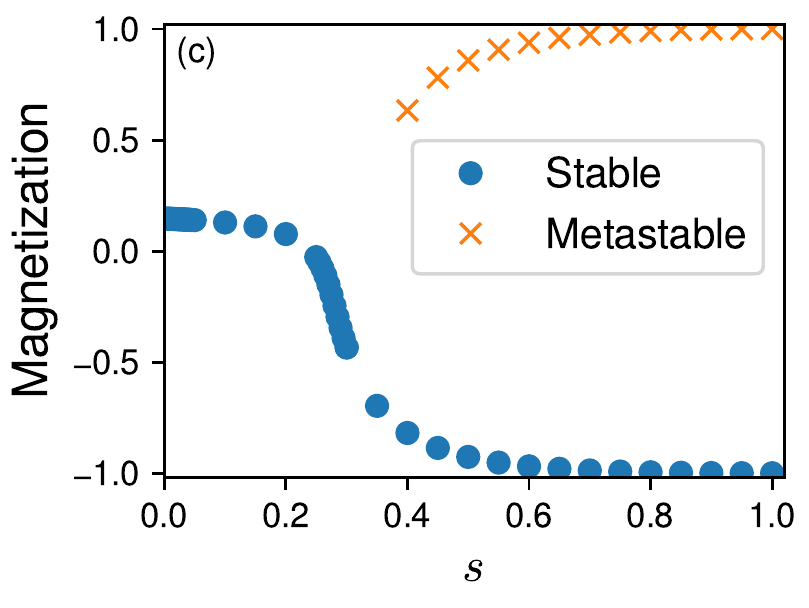}
  \caption{(Color online). Magnetization as a function of $s$ for each the reference point
  in Fig.~\ref{WP_fig:num_pt_positive_omega_negative_h}.
  The degrees of degeneracies are $g_{\text{u}} = 2$ and $g_{\text{l}} = 1$.
  The parameter $\omega$ is $0.42$ for all the figures.
  The longitudinal fields are (a) $h = -0.2$, (b) $h=-0.3$, and (c) $h=-0.4$.
  The blue dots represent the magnetization for stable states, and orange crosses the magnetization for metastable states.
  }
  \label{WP_fig:m_positive_omega_gu2gl1}
\end{figure}

Figure~\ref{WP_fig:m_positive_omega_gu2gl1} shows the magnetization as a function of $s$
for each the reference point in Fig.~\ref{WP_fig:num_pt_positive_omega_negative_h}.
The magnetization of the stable states changes at $s \simeq 0.4$ discontinuously
in Fig.~\ref{WP_fig:m_positive_omega_gu2gl1}(a),
meaning the system undergoes the single first-order phase transition.
Considering that the upper states are favored in the ground state of the quantum driver Hamiltonian,
and the lower states are favored in that of the final Hamiltonian,
the data indicate that the first-order phase transition results from the competition
between the initial and final Hamiltonians.
Meanwhile, the magnetization of the stable states for (b) and (c) varies continuously.
We thus considered that no first-order phase transitions occur for the cases (b) and (c).

\subsection{Case: $-1 < \omega < 0$ and $h\ge 0$}
\label{WP_sec:sub_lower-upper}

Let us show results for the case where the lower states are favored in the ground state of the quantum driver Hamiltonian,
whereas the upper states are favored in that of the final Hamiltonian.
Since the Perron--Frobenius theorem is not applicable to the case,
we cannot derive an equivalent quantum Ising model in the same way as the prior sections.
Accordingly, we investigated phase transition phenomena by using the numerical analysis.

\begin{figure}[tp]
  \centering
  \includegraphics{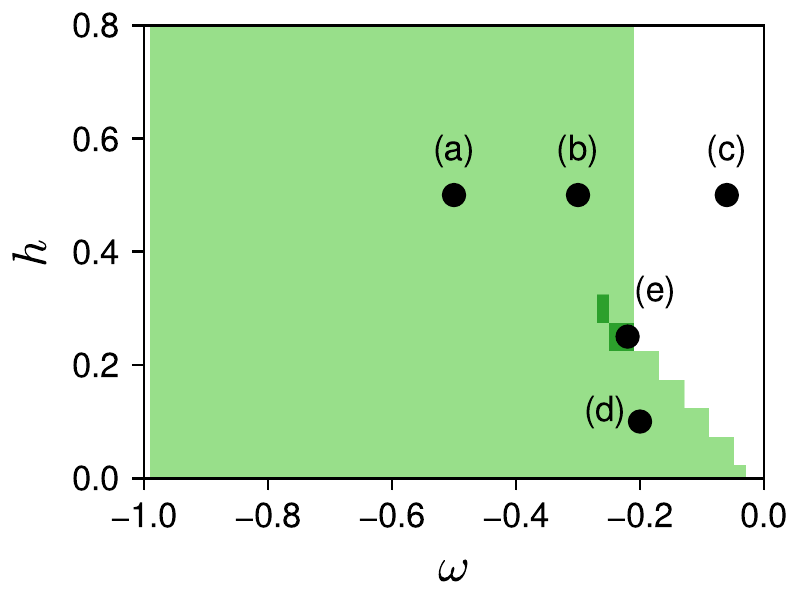}
  \caption{(Color online). The number of first-order phase transitions during a quantum annealing procedure
  in the fully-connected quantum Wajnflasz--Pick model for $-1 < \omega < 0$,  $h \ge 0$, $g_{\text{u}} = 3$, and $g_{\text{l}} = 1$.
  The light green region means that the system undergoes a first-order phase transition once in the procedure.
  A first-order phase transition occurs twice during the procedure in the dark green region.
  The dots represent reference points where we show the behavior of magnetization later.
  The coordinates of the reference points are (a) $(-0.5, 0.5)$, (b) $(-0.3, 0.5)$, (c) $(-0.06, 0.5)$,
  (d) $(-0.2, 0.1)$, and (e) $(-0.22, 0.25)$.
  }
  \label{WP_fig:num_pt_gu3gl1_negative_omega_positive_h}
\end{figure}

Figure~\ref{WP_fig:num_pt_gu3gl1_negative_omega_positive_h} shows the number of first-order phase transitions
during a quantum annealing procedure for $g_{\text{u}} = 3$ and $g_{\text{l}} = 1$.
A remarkable difference from the previous results for positive $\omega$ is that
the system undergoes a first-order phase transition twice under certain conditions (dark green region).
In addition, we describe that the cause of the first-order phase transitions in the light green region
differs from the previous results in this section.

\begin{figure}[tp]
  \centering
  \includegraphics[width=60mm]{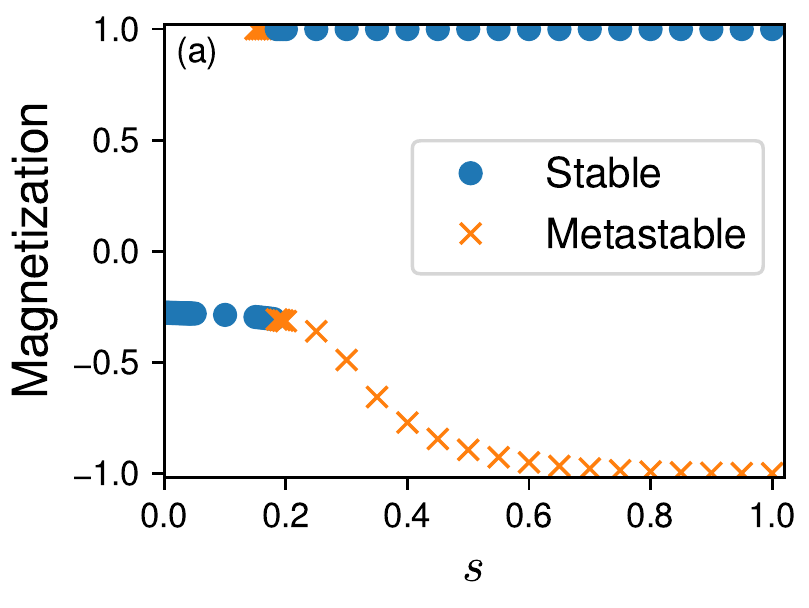}
  \includegraphics[width=60mm]{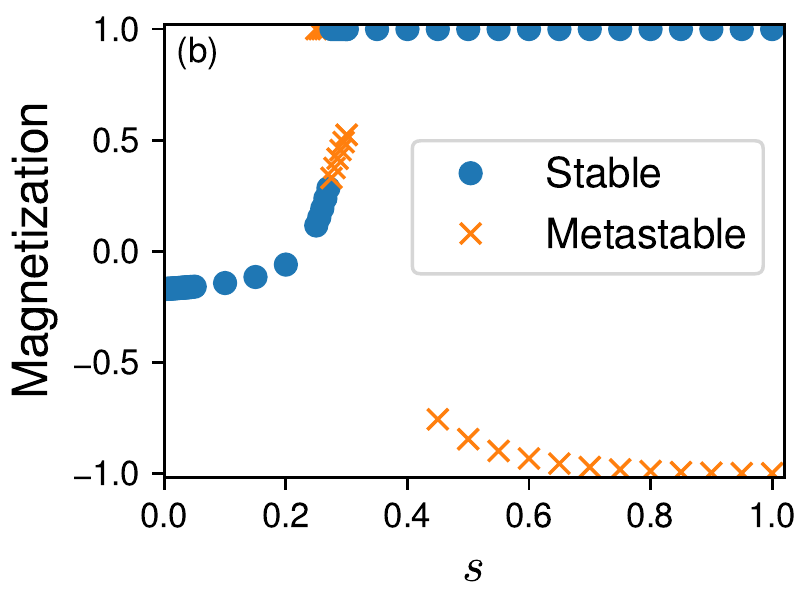}
  \includegraphics[width=60mm]{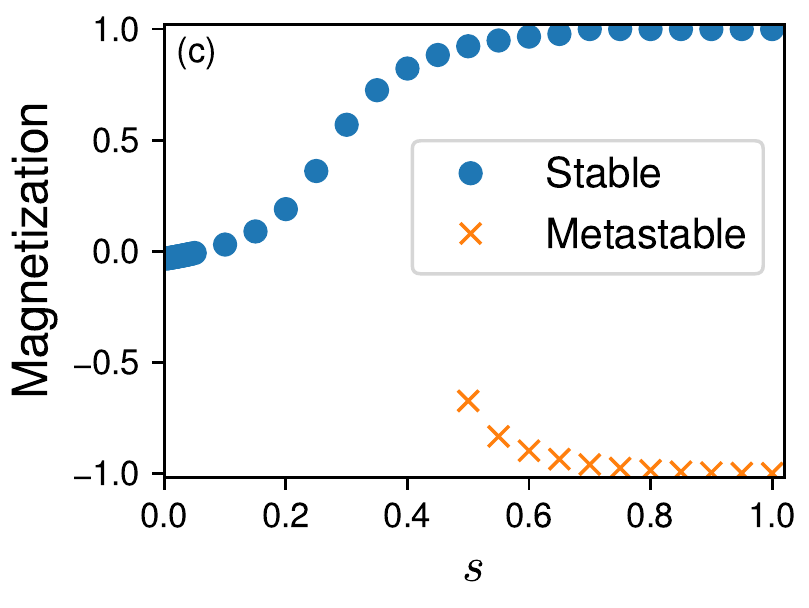}
  \includegraphics[width=60mm]{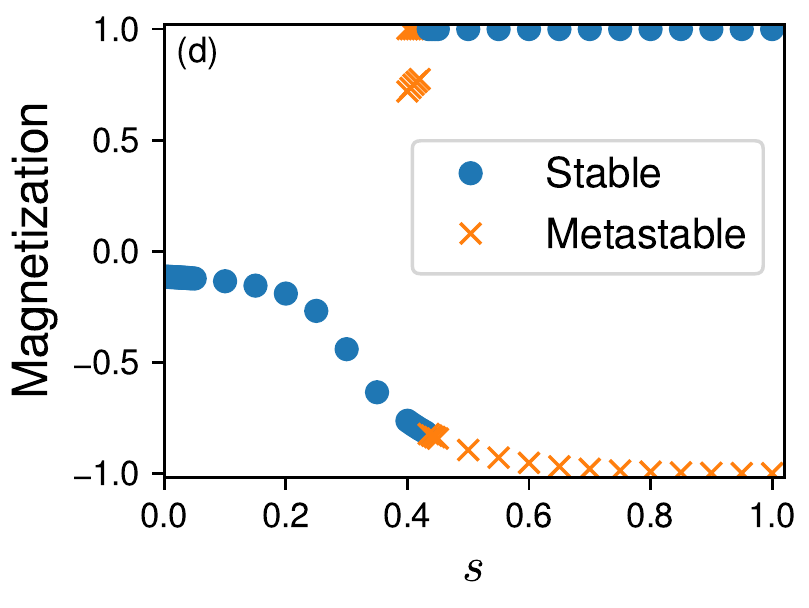}
  \caption{(Color online). Magnetization as a function of $s$ for each the reference point
  in Fig.~\ref{WP_fig:num_pt_gu3gl1_negative_omega_positive_h}.
  The degrees of degeneracies are $g_{\text{u}}= 3$ and $g_{\text{l}}= 1$.
  The parameters $(\omega, h)$ are (a) $(-0.5, 0.5)$, (b) $(-0.3, 0.5)$, (c) $(-0.06, 0.5)$,
  and (d) $(-0.2, 0.1)$.
  The blue dots represent the magnetization of stable states,
  and orange crosses the magnetization of metastable states.
  }
  \label{WP_fig:m_negative_omega_positive_h_gu3gl1_abcd}
\end{figure}

\begin{figure}[tp]
  \centering
  \includegraphics[width=70mm]{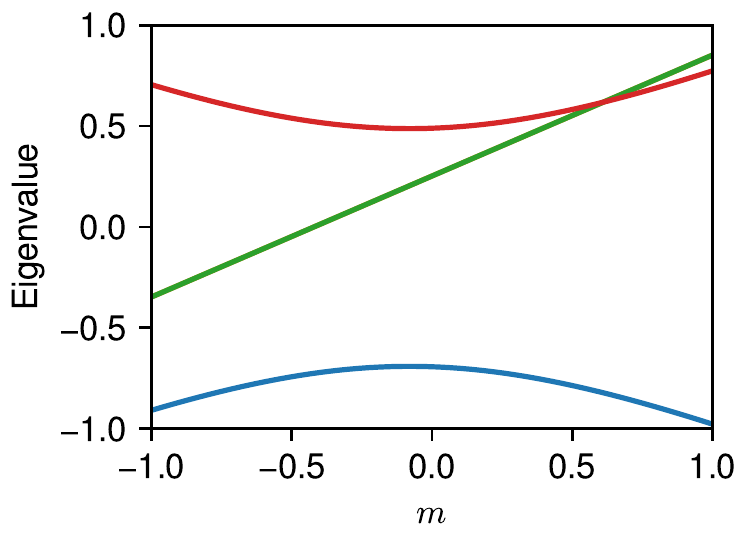}
  \caption{(Color online). The eigenvalues of the operator $\tilde{m}\hat{\tau}^{z} + (1-s)\hat{\tau}^{x}$
  as functions of $m$ for $g_{\text{u}} = 3$, $g_{\text{l}} = 1$, $s=0.3$, $\omega = -0.3$, and $h = 0.5$.
  Successive eigenvalues that can be fit by a continuous line are represented in the same color.
  The eigenvalues increasing linearly (green line) are degenerate.
  The maximum eigenvalue switches from a line of successive eigenvalues (red) to another line (green)
  at $m \simeq 0.6$.
  }
  \label{WP_fig:eigenvalues_operator_gu3gl1_s0p3_omega-0p3_h0p5}
\end{figure}

Figure~\ref{WP_fig:m_negative_omega_positive_h_gu3gl1_abcd} shows the magnetization as a function of $s$
for the reference points (a), (b), (c), and (d) in Fig.~\ref{WP_fig:num_pt_gu3gl1_negative_omega_positive_h}.
Although a first-order phase transition occurs at each of the reference points (a), (b), and (d),
the cause of the first-order phase transitions is different
from those described in the previous section~\ref{WP_sec:sub_upper-lower}.
Whereas the maximum eigenvalues $\lambda_\text{max}$ of the previous case $\omega \ge 0$
are denoted by the single continuous function of $m$ in Eq.~\eqref{WP_eq:lambda_max},
the maximum eigenvalue for $-1 < \omega < 0$ is represented by using multiple continuous functions
as shown in Fig.~\ref{WP_fig:eigenvalues_operator_gu3gl1_s0p3_omega-0p3_h0p5}.
Let us call such a continuous function a \textit{branch of eigenvalue}.
Since $\lambda_{\text{max}}$ is represented by a single branch of eigenvalue for $\omega \ge 0$,
the first-order phase transitions shown in the previous section~\ref{WP_sec:sub_upper-lower}
happens within the branch of eigenvalue.
On the other hand, the first-order phase transitions in Fig.~\ref{WP_fig:m_negative_omega_positive_h_gu3gl1_abcd}
occur between two branches of eigenvalue:
That is, magnetization of stable states before and after the first-order phase transition
in Fig.~\ref{WP_fig:m_negative_omega_positive_h_gu3gl1_abcd}(a) belongs to different branches of eigenvalue.
The magnetization corresponding to the linear branch of eigenvalue
in Fig.~\ref{WP_fig:eigenvalues_operator_gu3gl1_s0p3_omega-0p3_h0p5} (green line)
is $m = 1$, whereas the magnetization for the non-linear branch of eigenvalue (red line)
varies as $s$ changes.
Thus, we can distinguish the branch of eigenvalue from the behavior of magnetization.
The first-order phase transitions in Fig.~\ref{WP_fig:m_negative_omega_positive_h_gu3gl1_abcd}(b)
and Fig.~\ref{WP_fig:m_negative_omega_positive_h_gu3gl1_abcd}(d) occur
between the linear and the non-linear branches of eigenvalues too.
The sequence of points of magnetization $m = 1.0$ belongs to the linear branch of eigenvalue, and the other points
to the non-linear branch of eigenvalue.
A difference from the result for the reference point (a) is that there exist unreachable metastable states
in Fig.~\ref{WP_fig:m_negative_omega_positive_h_gu3gl1_abcd}(b)
and Fig.~\ref{WP_fig:m_negative_omega_positive_h_gu3gl1_abcd}(d).
In Fig.~\ref{WP_fig:m_negative_omega_positive_h_gu3gl1_abcd}(b),
the metastable states with the smallest magnetization belonging to the non-linear branch of eigenvalue
are not reachable from the initial stable state.
The metastable states with larger magnetization belonging to the non-linear branch of eigenvalue
are unreachable as shown in Fig.~\ref{WP_fig:m_negative_omega_positive_h_gu3gl1_abcd}(d).
We did not observe first-order phase transitions for $\omega = -0.06$ and $h = 0.5$
(Fig.~\ref{WP_fig:m_negative_omega_positive_h_gu3gl1_abcd}(c))
with our numerical method.

Limitation of our numerical method should be noted.
Our numerical method cannot detect first-order phase transitions
where a magnetization jump is smaller than the threshold $0.3$.
We confirmed that the magnetization jumps between the linear and the non-linear branches of eigenvalue become smaller
as $\omega$ increases.
Hence, our numerical method probably failed to detect first-order phase transitions
in the white region in Fig.~\ref{WP_fig:num_pt_gu3gl1_negative_omega_positive_h}.
We should also mention that our numerical method cannot find first-order phase transitions
that happens at a point $s$ smaller than $0.001$.
We observed that points of first-order phase transition between the linear and the non-linear branches of eigenvalue
get smaller as the longitudinal field $h$ increases, or the parameter $\omega$ decreases, or both.
Accordingly, the detection of first-order phase transitions is difficult for such parameters.

\begin{figure}[tp]
  \centering
  \includegraphics[width=70mm]{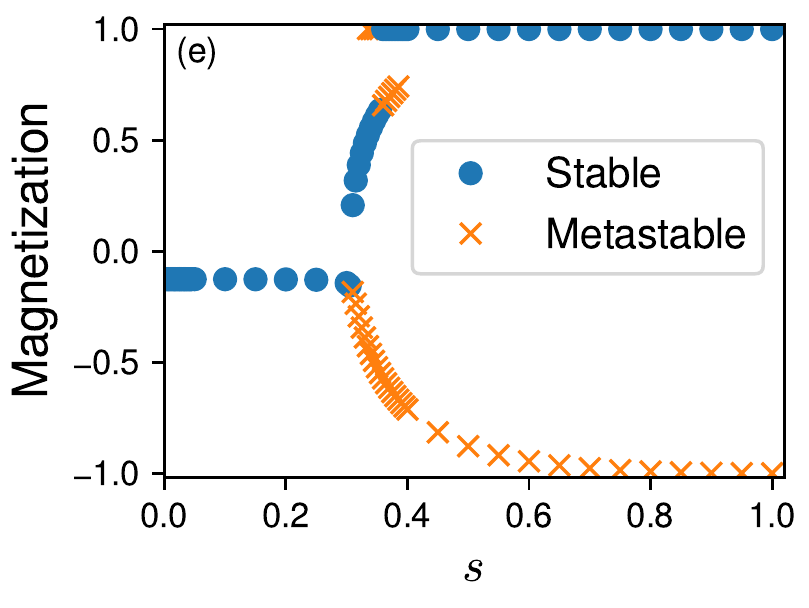}
  \caption{(Color online). Magnetization as a function of $s$ for the reference point (e)
  in Fig.~\ref{WP_fig:num_pt_gu3gl1_negative_omega_positive_h}.
  The degrees of degeneracies are $g_{\text{u}} = 3$ and $g_{\text{l}} = 1$.
  The value of longitudinal field is $h=0.25$,
  and the values of parameter $\omega$ is $-0.22$.
  The blue dots represent the magnetization of stable states,
  and orange crosses the magnetization of metastable states.
  }
  \label{WP_fig:m_negative_omega_positive_h_gu3gl1_e}
\end{figure}

A characteristic behavior that a first-order phase transition occurs twice
is observed in a small area near the reference point (e) in Fig.~\ref{WP_fig:num_pt_gu3gl1_negative_omega_positive_h}.
We show the magnetization for the point in Fig.~\ref{WP_fig:m_negative_omega_positive_h_gu3gl1_e}.
We found that the system undergoes a first-order phase transition within the non-linear branch of eigenvalue
followed by another first-order phase transition between the linear and the non-linear branches of eigenvalue.
Although we detected the characteristic behavior in the small region,
we consider that the behavior can be observed also in the lower right of the light green region
in Fig.~\ref{WP_fig:num_pt_gu3gl1_negative_omega_positive_h}
because of the aforementioned limitation of our numerical method arisen from the threshold value.
We obtained similar results for $g_{\text{u}} = 2, 4, 5, 6$ and $g_{\text{l}}= 1$.

\begin{figure}[tp]
  \centering
  \includegraphics[width=70mm]{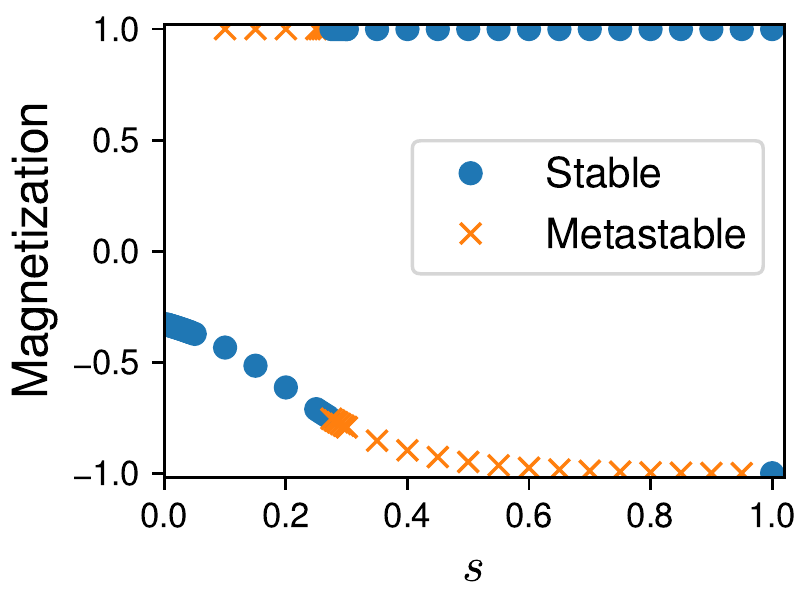}
  \includegraphics[width=70mm]{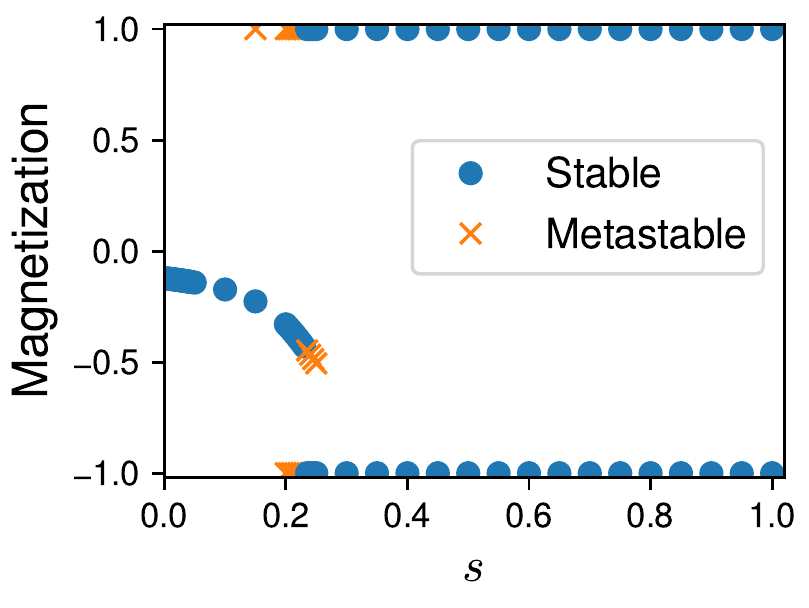}
  \caption{(Color online). Magnetization as a function of $s$ for $g_{\text{u}} = 3$ and $g_{\text{l}} = 1$ (top),
  and $g_{\text{u}} = 3$ and $g_{\text{l}} = 2$ (bottom) in the absence of a longitudinal field.
  The values of parameter $\omega$ is $-0.6$.
  The blue dots represent the magnetization of stable states,
  and orange crosses the magnetization of metastable states.
  }
  \label{WP_fig:m_negative_omega_h0}
\end{figure}

Let us next consider the case with $g_{\text{l}} > 1$.
A difference from the case with $g_{\text{l}} = 1$ is that another branch of eigenvalue appears
because of the degenerate lower level.
The eigenvalue of the new branch scales linearly as a function of $m$, and corresponding magnetization is $m = -1$.
We observed that minima of the pseudo-free energy corresponding to the linear branch of eigenvalue with $m = -1$
are always metastable states for $h > 0$.
Hence, phase transition phenomena are similar to the case with $g_{\text{l}} = 1$ when $h > 0$.
In the absence of the longitudinal field, the behavior of magnetization differs from the case with $g_{\text{l}} = 1$.
Figure~\ref{WP_fig:m_negative_omega_h0} shows the magnetization as a function of $s$
for $g_{\text{u}} = 3$ and $g_{\text{l}} = 1$ (top),
and $g_{\text{u}} = 3$ and $g_{\text{l}} = 2$ (bottom).
The parameter $\omega$ is $-0.6$.
Both of the minima of the pseudo-free energy corresponding to the linear branches of eigenvalue
are the stable states for $g_{\text{u}} = 3$ and $g_{\text{l}} = 2$
after the first-order phase transition,
whereas only the minimum of the pseudo-free energy corresponding to the linear branch of eigenvalue with $m=1$
is the stable state for $g_{\text{u}}=3$ and $g_{\text{l}}=1$.

\subsection{Case: $-1 < \omega < 0$ and $h<0$}
\label{WP_sec:sub_lower-lower}

We show results where the lower states are favored in both of the ground states of the quantum driver Hamiltonian
and the final Hamiltonian.
Behaviors described below can be understood through the same discussion of the previous sections.

\begin{figure}[tp]
  \centering
  \includegraphics{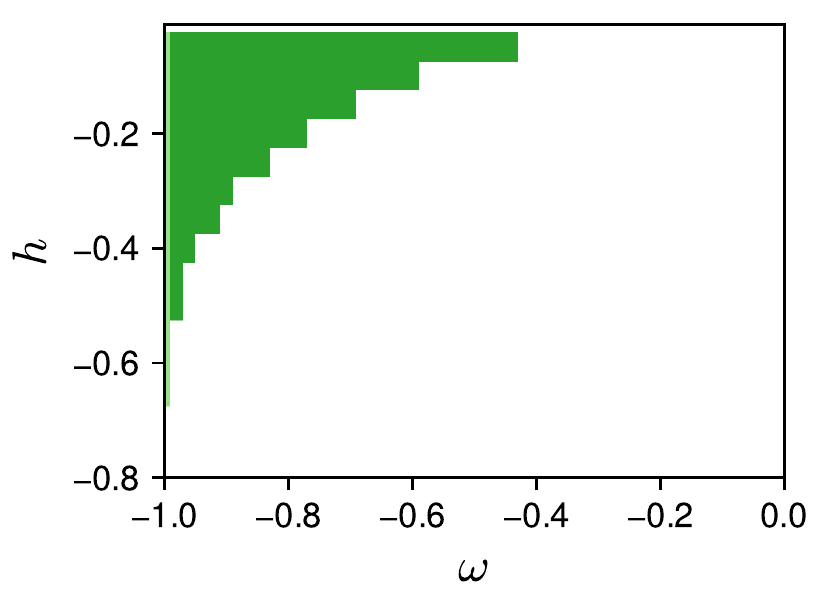}
  \caption{(Color online). The number of first-order phase transitions during a quantum annealing procedure
  in the fully-connected quantum Wajnflasz--Pick model for $-1 < \omega < 0$,  $h < 0$, $g_{\text{u}} = 2$, and $g_{\text{l}} = 1$.
  A first-order phase transition occurs once (light green region), and twice (dark green region).
  }
  \label{WP_fig:num_pt_gu2gl1_negative_omega_negative_h}
\end{figure}

\begin{figure}[tp]
  \centering
  \includegraphics[width=70mm]{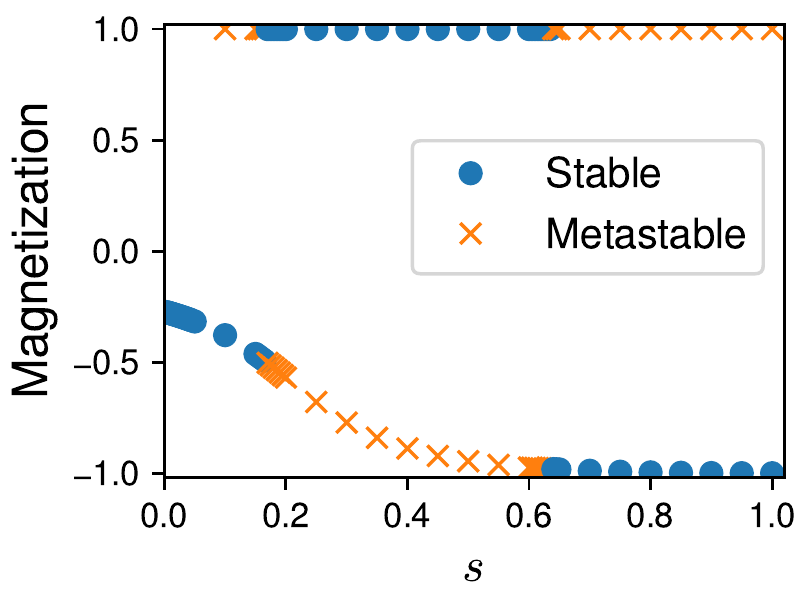}
  \caption{(Color online). Magnetization as a function of $s$ for $g_{\text{u}} = 2$ and $g_{\text{l}} = 1$.
  The values of parameters are $\omega = -0.8$ and $h = -0.1$.
  The blue dots represent the magnetization of stable states,
  and orange crosses the magnetization of metastable states.
  }
  \label{WP_fig:m_negative_omega_negative_h_gu2gl1}
\end{figure}

We show the number of first-order phase transitions occurring a quantum annealing procedure
for $g_{\text{u}} = 2$ and $g_{\text{l}} = 1$ in Fig.~\ref{WP_fig:num_pt_gu2gl1_negative_omega_negative_h}.
We observed that the system undergoes a first-order phase transition twice in the upper left corner of the figure.
We consider that the light green region becomes dark green region if we investigate magnetization for further smaller $s$,
because, as mentioned above, our numerical method cannot detect first-order phase transitions
that occur at a very small value of $s$.
We confirmed that the first point $s$ of the first-order phase transition decreases
as $\omega$ decreases.
Figure~\ref{WP_fig:m_negative_omega_negative_h_gu2gl1} shows the behavior of magnetization
where a first-order phase transition occurs twice.
Although the stable states have negative magnetization at the beginning and the end of quantum annealing procedure
due to the ground states of the quantum driver Hamiltonian and the final Hamiltonian,
the magnetization belonging to the linear branch of eigenvalue with $m = 1$ becomes the stable states
in the intermediate values of $s$.
The successive first-order phase transitions are caused by transitions of the stable state
between the linear and the non-linear branches of eigenvalue.

\begin{figure}[tp]
  \centering
  \includegraphics[width=70mm]{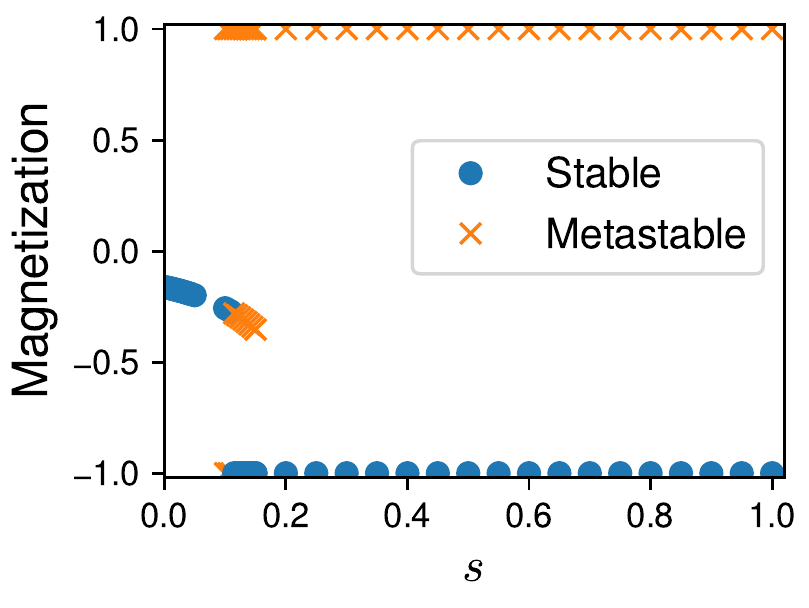}
  \caption{(Color online). Magnetization as a function of $s$ for $g_{\text{u}} = 3$ and $g_{\text{l}} = 2$.
  The values of parameters are $\omega = -0.8$ and $h = -0.1$.
  The blue dots represent the magnetization of stable states,
  and orange crosses the magnetization of metastable states.
  }
  \label{WP_fig:m_negative_omega_negative_h_gu3gl2}
\end{figure}

We did not observe such successive first-order phase transitions for $g_{\text{l}} > 1$.
Figure~\ref{WP_fig:m_negative_omega_negative_h_gu3gl2} represents the magnetization
as a function of $s$ for $g_{\text{u}} = 3$, $g_{\text{l}} = 2$, $\omega = -0.8$, and $h = -0.1$.
Since the linear branch of eigenvalue with $m = -1$ is more stable than that with $m = 1$,
a first-order phase transition to the branch of eigenvalue with $m = 1$ does not occur.
Consequently, the single first-order phase transition occurs.

\subsection{Results for balanced degeneracies $g_{\text{u}} = g_{\text{l}}$}
\label{WP_sec:sub_results_for_balanced}

Let us focus on the balanced case $g_{\text{u}} = g_{\text{l}}$.
We assume that $h \ge 0$ without loss of generality.
We do not discuss about the case $g_{\text{u}} = g_{\text{l}} = 1$,
since the system is completely equivalent to the usual quantum Ising model.

First, we consider the case with $\omega \ge 0$.
As discussed in Sec.~\ref{WP_sec:sub_upper-upper},
we can derive the equivalent quantum Ising model.
From Eq.~\eqref{WP_eq:quantum_heff}, the effective longitudinal field $h_{\text{eff}}$
coincides with the original longitudinal field $h$.
In other words, quantum annealing paths are parallel to the $s$ axis.
Consequently, we conclude that the system undergoes a second-order phase transition when $h = 0$,
and no phase transitions when $h > 0$.

\begin{figure}[tp]
  \centering
  \includegraphics[width=70mm]{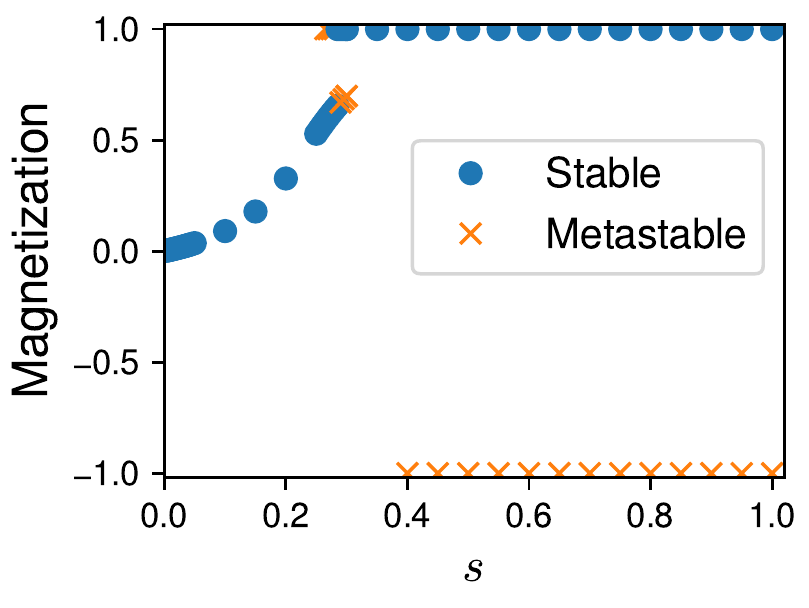}
  \includegraphics[width=70mm]{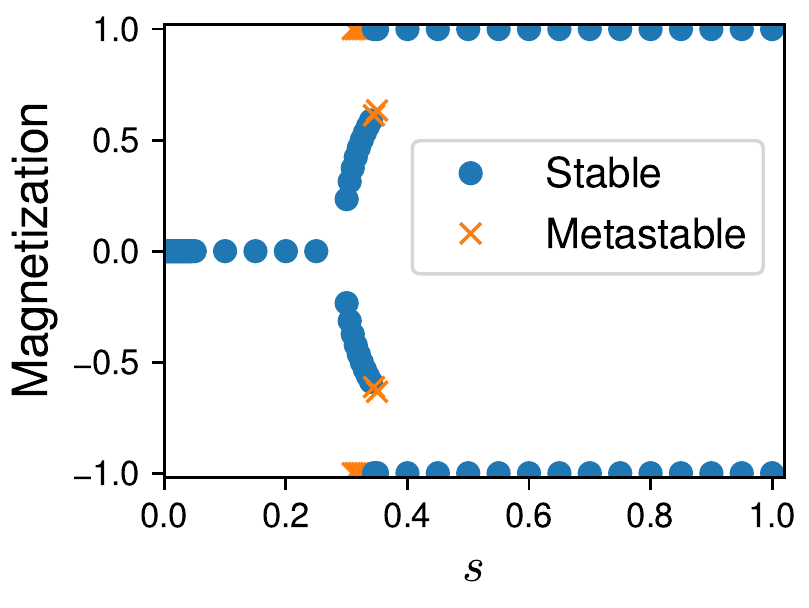}
  \caption{(Color online). Magnetization as a function of $s$ for $g_{\text{u}} = 2$ and $g_{\text{l}} = 2$.
  The values of parameters are $\omega = -0.4$ and $h = 0.5$ (top), and $\omega = -0.4$ and $h = 0.0$ (bottom).
  The blue dots represent the magnetization of stable states,
  and orange crosses the magnetization of metastable states.
  }
  \label{WP_fig:m_negative_omega_positive_h_gu2gl2}
\end{figure}

Next, we show results for $-1 < \omega < 0$.
Figure~\ref{WP_fig:m_negative_omega_positive_h_gu2gl2} represents magnetization as a function of $s$
for $g_{\text{u}} = g_{\text{l}} = 2$, $\omega = -0.4$ and $h = 0.5$ (top), and $\omega = -0.4$ and $h = 0.0$ (bottom).
For $h = 0.5$, the magnetization belonging to the linear branch of eigenvalue with $m = 1$
is more stable than that with $m = -1$.
We observed the first-order phase transition from the non-linear branch of eigenvalue to the linear branch of eigenvalue.
Even when a longitudinal field is absent, we observed the first-order phase transition
between the linear and the non-linear branches of eigenvalue.

\section{Conclusion}
\label{sec:conclusion}

We studied the nature of phase transitions in the fully-connected quantum Wajnflasz--Pick model
in order to reveal the effect of the degeneracy of the upper and the lower levels
and the state transitions within each of the upper and the lower states.
We analyzed the system whose Hamiltonian is given as Eq.~\eqref{WP_eq:H_QA} in accordance with the practice of QA and AQC.
Since we considered the quantum Wajnflasz--Pick model consisting of fully-connected variables,
we used the mean-field analysis to calculate the magnetization of the system.

We classified our results for unbalanced case $g_{\text{u}} > g_{\text{l}}$ into four groups
according to whether the upper states are favored or not
in the ground states of the initial Hamiltonian and the final Hamiltonian.

We first considered the case $\omega \ge 0$ and $h \ge 0$, where the upper states are favored
in both the initial and the final Hamiltonian.
We derived the equivalent quantum Ising model that has an effective longitudinal field
depending on $s$.
The result means that quantum annealing path on the $(s, h)$ plane is not necessarily parallel with the $s$ axis
even though we fix the longitudinal field $h$.
We showed that no first-order phase transitions happen in this case.

Second, we showed the results for $\omega \ge 0$ and $h < 0$, where the upper states are favored in the initial ground state,
whereas the lower states are favored in the final ground state.
We applied the same method as the previous case.
As a result, we found that a single first-order phase transition occurs
when the annealing path traverses the first-order phase boundary.
of the quantum Ising model.
We considered that the first-order phase transition is caused
by the competition between the initial and the final ground states.

Third, we discussed about the case $-1 < \omega < 0$ and $h \ge 0$, where the lower states are favored in the initial ground state, and the upper states are favored in the final ground state.
We observed the first-order phase transitions between the linear and the non-linear branches of eigenvalue
that we defined in Sec.~\ref{WP_sec:sub_lower-upper}.
This result indicates that the cause of the first-order phase transitions is not only the competition
between the initial and the final ground states.
but also the existence of the branches of eigenvalue that do not appear in the usual quantum Ising model.
We observed characteristic phenomena that the system undergoes the first-order phase transition
within the non-linear branch of eigenvalue as well as the transition
between the linear and the non-linear branches of eigenvalue.
Owing to the first-order phase transitions between the linear and the non-linear branches of eigenvalue,
unreachable metastable states appear under certain values of longitudinal field,
which is also a characteristic behavior compared to the usual quantum Ising model.

Next, we considered the case $-1 < \omega < 0$ and $h < 0$.
Although the lower states are favored in both the initial and the final ground states,
we observed the region where magnetization of the stable states become unity in the intermediate values of $s$.
We showed that the first-order phase transition between the linear and the non-linear branches of eigenvalue occurs twice.

Finally, we considered the balanced case $g_{\text{u}} = g_{\text{l}}$.
We showed that the system is completely equivalent to the quantum Ising model when $\omega \ge 0$.
For $-1 < \omega < 0$, we observed the first-order phase transitions between the linear and the non-linear branches of eigenvalue,
which is similar to the unbalanced case with $-1 < \omega < 0$ and $h > 0$.

In short, the phase transition phenomena in the fully-connected quantum Wajnflasz--Pick model with $\omega \ge 0$
can be explained by the usual quantum Ising model.
For $\omega \ge 0$, the phase transitions happen within the non-linear branch of eigenvalue.
For $-1 < \omega < 0$, the linear branches of eigenvalue appears because of the degenerate
upper and lower levels.
We observed the first-order phase transitions between the linear and the non-linear branches of eigenvalue,
which do not happen in the usual quantum Ising model.

In the present paper, we investigated phase transition phenomena
in the fully-connected quantum Wajnflasz--Pick model.
We revealed that the fully-connected quantum Wajnflasz--Pick model undergoes the characteristic first-order phase transitions
resulting from the degenerate upper and lower levels.
Since the fully-connected quantum Wajnflasz--Pick model undergoes the first-order phase transitions
that do not happen in the usual quantum Ising model,
we concluded that using the two-level system with the degenerate upper and lower levels as a qubit
exacerbates the performance of AQC.
Nevertheless, the fully-connected quantum Wajnflasz--Pick model is attractive
from the viewpoint of condensed matter physics.
The model undergoes the first-order phase transitions between the linear and the non-linear branches
of eigenvalue, which do not appear in the usual quantum Ising model.
Since the fully-connected quantum Wajnflasz--Pick model has the isolated metastable states,
devices based on the model have potential applications for switches.

\begin{acknowledgment}
This paper is partly based on results obtained from a project commissioned by the New Energy and Industrial Technology Development Organization (NEDO), Japan.
One of the authors (S.~T.) was supported by JST, PRESTO Grant Number JPMJPR1665, Japan and JSPS KAKENHI Grant Numbers 15K17720, 15H03699.
One of the author (S.~K.) was supported by JST, CUPAL.
\end{acknowledgment}

\appendix

\section{Derivation of pseudo-free energy}
\label{appdx:pseudo-free-energy}
We consider the Hamiltonian given by Eq.~\eqref{WP_eq:quantum_WP}.
The partition function for the Hamiltonian can be expressed as
\begin{align}
  Z &= \lim_{M \to \infty} Z_{M} \notag \\
    &= \lim_{M \to \infty} \sum_{\{ \tau_{i}^{z} \}}
      \langle \{ \tau_{i}^{z} \} |
      \Biggl(
      \exp \Biggl\{
      \frac{s\beta N}{M} \Biggl[\Biggl(\frac{1}{N} \sum_{i=1}^{N}\hat{\tau}_{i}^{z}\Biggr)^{2}
      + \frac{h}{N}\sum_{i=1}^{N}\hat{\tau}_{i}^{z}\Biggr]
      \Biggr\}
      \notag \\
    &\quad\times
      \exp \Biggl\{ \frac{(1-s)\beta}{M}\sum_{i=1}^{N}\hat{\tau}_{i}^{x} \Biggr\}
      \Biggr)^{M}| \{\tau_{i}^{z}\}\rangle
\end{align}
by using the Suzuki--Trotter decomposition~\cite{suzuki1976relationship}.
Here, $M$ is an integer called the Trotter number, and $\beta$ represents an inverse temperature.
The summation $\sum_{ \{ \tau_{i}^{z} \}}$ is taken over all the eigenstates $\lvert \{ \tau_{i}^{z} \} \rangle$
of $\bigotimes_{i=1}^{N}\hat{\tau}_{i}^{z}$.
To translate the system to a classical system, we put the following closure relations
\begin{align}
  \hat{1}(\alpha) &= \sum_{ \{ \tau_{i}^{z} \}} | \{ \tau_{i}^{z} \} \rangle\langle \{ \tau_{i}^{z} \} |
                    \times
                    \sum_{ \{ \tau_{i}^{x} \}} | \{ \tau_{i}^{x} \} \rangle\langle \{ \tau_{i}^{x} \} |
\end{align}
just before the $\alpha$th exponential operator including $\hat{\tau}^{x}$.
Here, $| \{ \tau_{i}^{x} \}\rangle$ represents an eigenstate of $\bigotimes_{i=1}^{N}\hat{\tau}_{i}^{x}$,
and $\sum_{ \{ \tau_{i}^{x} \}}$ the summation over all the eigenstates.
The partition function for given $M$ reads
\begin{align}
  Z_{M} &= \sum_{ \{ \tau_{i}^{z} \}}\sum_{ \{ \tau_{i}^{x} \}}\prod_{\alpha = 1}^{M}
          \exp \biggl\{
          \frac{\beta s N}{M} \biggl[
          \biggl(\frac{1}{N} \sum_{i}\tau_{i}^{z}(\alpha)\biggr)^{2} + \frac{h}{N}\sum_{i}\tau_{i}^{z}(\alpha)
          \biggr] \notag \\
        & + \frac{\beta (1-s)}{M} \sum_{i}\tau_{i}^{x}(\alpha)
          \biggr\}
          \times
          \prod_{i}\langle \tau_{i}^{z}(\alpha) | \tau_{i}^{x}(\alpha) \rangle
          \langle \tau_{i}^{x}(\alpha) | \tau_{i}^{z}(\alpha + 1) \rangle ,
  \label{WP_eq:apdx_ZM}
\end{align}
where $| \tau_{i}^{z}(1) \rangle = | \tau_{i}^{z}(M + 1)\rangle$ for all $i$.
We introduce an integral representation of a delta function:
\begin{align}
  &\delta\biggl(Nm(\alpha) - \sum_{i}\tau_{i}^{z}(\alpha)\biggr) \notag \\
  &= \int d \tilde{m}(\alpha) \exp \biggl[-\tilde{m}(\alpha)\frac{\beta}{M}
    \biggl(Nm(\alpha) - \sum_{i} \tau_{i}^{z}(\alpha)\biggr)\biggr]
\end{align}
to linearize the quadratic term in Eq.~\eqref{WP_eq:apdx_ZM}.
Here, $m$ represents an order parameter corresponding to magnetization, and $\tilde{m}$ is the conjugate variable
of $m$.
The partition function $Z_{M}$ is rewritten as
\begin{align}
  Z_{M}
  &=
    \sum_{ \{ \tau_{i}^{z} \}}\sum_{ \{ \tau_{i}^{x} \}}
    \idotsint \prod_{\alpha} dm(\alpha) d \tilde{m}(\alpha) \notag \\
  &\quad\times 
    \exp \biggl\{ \frac{\beta N}{M} \bigl[ s\bigl([m(\alpha)]^{2} + h m(\alpha)\bigr) - \tilde{m}(\alpha)m(\alpha)\bigr ] \biggr\}
    \notag \\
  &\quad\times
    \prod_{i} \exp \Bigl\{ \frac{\beta}{M} (\tilde{m}(\alpha)\tau_{i}^{z}(\alpha)
    + (1-s)\tau_{i}^{x}(\alpha)) \Bigr\}\notag \\
  &\quad\times
    \langle \tau_{i}^{z}(\alpha) | \tau_{i}^{x}(\alpha) \rangle
    \langle \tau_{i}^{x}(\alpha) | \tau_{i}^{z}(\alpha + 1) \rangle .
    \notag \\
  &=
    \idotsint \prod_{\alpha} dm(\alpha) d \tilde{m}(\alpha)\notag \\
  &\quad\times
    \exp \biggl\{ \frac{\beta N}{M} \bigl[s\bigl([m(\alpha)]^{2} + h m(\alpha)\bigr) - \tilde{m}(\alpha)m(\alpha)\bigr] \biggr\}
    \notag \\
  &\quad\times
    \exp \Biggl\{ N \ln \Tr \prod_{\alpha}
    \exp \biggl\{
    \frac{\beta}{M} (\tilde{m}(\alpha)\tau^{z}(\alpha)
    + (1-s)\tau^{x}(\alpha)) \biggr\} \notag \\
  &\quad \times
    \langle \tau^{z}(\alpha) | \tau^{x}(\alpha) \rangle
    \langle \tau^{x}(\alpha) | \tau^{z}(\alpha + 1) \rangle
    \Biggr\}.
\end{align}
Here, $\Tr$ denotes the summation over all the possible values of $\tau^{z}(\alpha)$ and $\tau^{z}(\alpha)$.
The saddle-point conditions with regard to the order parameters $\{ m(\alpha) \}$ leads
\begin{align}
  \tilde{m}(\alpha) &= s(2m(\alpha) + h).
\end{align}
Applying the static ansatz that removes $\alpha$ dependence of order parameters, $m(\alpha) \equiv m$,
and the inverse operation of the Suzuki--Trotter decomposition, we have
\begin{align}
  Z_{M} &= \int dm\, \exp \biggl\{ -N \beta
          \biggl[s m^{2} - \frac{1}{\beta} \ln \Tr \exp \beta (\tilde{m} \hat{\tau}^{z} + (1-s) \hat{\tau}^{x})\biggr]
          \biggr\}.
\end{align}
The pseudo-free energy can be obtained by using the saddle point condition again:
\begin{align}
  f &= s m^{2} - \frac{1}{\beta} \ln \Tr \exp \beta (\tilde{m} \hat{\tau}^{z} + (1-s) \hat{\tau}^{x})
      \label{WP_eq:apdx_pf}
\end{align}
In the low-temperature limit $\beta \to \infty$, only the largest eigenvalue of the operator $\tilde{m} \hat{\tau}^{z} + (1-s) \hat{\tau}^{x}$ contributes:
\begin{align}
  f &= s m^{2} - \lambda_{\text{max}},
      \label{WP_eq:apdx_pf_low_temp}
\end{align}
where $\lambda_{\text{max}}$ denotes the largest eigenvalue.

If $\omega \ge 0$, the largest eigenvalue is evaluated as follows.
The Perron-Frobenius theorem ensures that the eigenvector for the largest eigenvalue of the operator
$\tilde{m} \hat{\tau}^{z} + (1-s) \hat{\tau}^{x}$
is unique and the elements of the eigenvector are positive.
Taking into account the symmetry of the operator with regards to reordering of the basis,
the eigenvector can be expressed as a vector consisting of two sections.
The first section includes the same $g_{\text{u}}$ elements, and the other section includes the same $g_{\text{l}}$ elements:
\begin{align}
  \boldsymbol{v} &= (\underbrace{v_{\text{u}}, \dotsc , v_{\text{u}}}_{g_{\text{u}}},
           \underbrace{v_{\text{l}}, \dotsc , v_{\text{l}}}_{g_{\text{l}}})^{\top},
\end{align}
where $v_{\text{u}}$ and $v_{\text{l}}$ are real and positive.
Hence, the eigenvalue equation is reduced to equations with two unknowns $v_{\text{p}}$ and $v_{\text{m}}$.
Solving the equations, we have the maximum eigenvalue of the operator:
\begin{align}
  \lambda_{\text{max}} &= \frac{1}{2} \left\{ (g_{\text{u}} + g_{\text{l}} - 2)\frac{(1-s)}{c}\omega \right.\notag \\
                       &+\left.
                         \sqrt{\left\{ (g_{\text{u}} - g_{\text{l}})\frac{1-s}{c}\omega + 2 \tilde{m} \right\}^{2}
                         + 4 g_{\text{u}}g_{\text{l}}\left(\frac{1-s}{c}\right)^{2}}\right\}.
  \label{WP_eq:apdx_positive_case_largest_eig}
\end{align}
Here, $c$ is the normalization factor of the operator~\eqref{WP_eq:def_taux}.
Consequently, the pseudo-free energy in the low-temperature limit is given by Eq.~\eqref{WP_eq:apdx_pf_low_temp}
with the largest eigenvalue~\eqref{WP_eq:apdx_positive_case_largest_eig}.

\section{Derivation of equivalent quantum Ising model}
\label{appdx:equivalent_spin-1/2}

In this section, we derive a Hamiltonian of quantum Ising model
whose pseudo-free energy is equivalent to that of the fully-connected quantum Wajnflasz--Pick model for $\omega \ge 0$.

First, we derive the pseudo-free energy of the following Hamiltonian of quantum Ising model:
\begin{align}
  \hat{H} = s\left[-\frac{1}{N}\sum_{i,j = 1}^{N} \hat{\sigma}_{i}^{z}\hat{\sigma}_{j}^{z}
  -h \sum_{i=1}^{N} \hat{\sigma}_{i}^{z}\right]
  - k(1-s)\sum_{i=1}^{N} \hat{\sigma}_{i}^{x}.
  \label{WP_eq:apdx_H_spin-1/2}
\end{align}
Here, $\hat{\sigma}_{i}^{z}$ denotes the $z$ component of the Pauli matrices acting on a site $i$,
and $\hat{\sigma}_{i}^{x}$ denotes the $x$ component.
The parameter $k$ is introduced for convenience sake.
Recall that the fully-connected quantum Wajnflasz--Pick model is a generalization of quantum Ising model.
We can reduce the Hamiltonian of the fully-connected quantum Wajnflasz--Pick model~\eqref{WP_eq:quantum_WP}
to the Hamiltonian of quantum Ising model~\eqref{WP_eq:apdx_H_spin-1/2}
by using $g_{\text{u}} = g_{\text{l}} = 1$, and replacing the factor $(1-s)$ by $k(1-s)$.
Here, we used the fact that the spectral norms of $\hat{\sigma}^{z}$ and $\hat{\sigma}^{x}$ are equal,
so that the normalization factor is unity, $c = 1$.
Consequently, the pseudo-free energy is given from
Eqs.~\eqref{WP_eq:apdx_pf_low_temp} and \eqref{WP_eq:apdx_positive_case_largest_eig}
as
\begin{align}
  f = sm^{2} - \sqrt{\tilde{m}^{2} + k^{2}(1-s)^{2}}.
  \label{WP_eq:apdx_f_spin-1/2}
\end{align}
Here, $\tilde{m} = s(2m+h)$.

Next, let us derive a Hamiltonian of quantum Ising model
whose pseudo-free energy is the same as
Eqs.~\eqref{WP_eq:apdx_pf_low_temp} and \eqref{WP_eq:apdx_positive_case_largest_eig}.
Once $h$ in Eq.~\eqref{WP_eq:apdx_H_spin-1/2} is replaced with $h + (1-s)(g_{\text{u}} - g_{\text{l}})\omega / 2 s c$,
and $k$ with $\sqrt{g_{\text{u}}g_{\text{l}}}/c$,
Eq.~\eqref{WP_eq:apdx_f_spin-1/2} reads
\begin{align}
  f = sm^{2} - \sqrt{\left\{\tilde{m} + (g_{\text{u}} - g_{\text{l}})\frac{1-s}{2c}\omega\right\}^{2}
  + g_{\text{u}}g_{\text{l}}\left(\frac{1-s}{c}\right)^{2}}.
  \label{WP_eq:apdx_f_spin-1/2_equivalent}
\end{align}
The pseudo-free energy~\eqref{WP_eq:apdx_f_spin-1/2_equivalent} is the same
as the pseudo-free energy for the fully-connected quantum Wajnflasz--Pick model given by
Eqs.~\eqref{WP_eq:apdx_pf_low_temp} and \eqref{WP_eq:apdx_positive_case_largest_eig}
except for the term independent of $m$.
Since the term does not change the minima of the pseudo-free energy,
the term is negligible when calculating magnetization.
As a result, we obtain the equivalent quantum Ising model,
\begin{align}
  \hat{H}' = s\left[ -\frac{1}{N}\sum_{i,j = 1}^{N}\hat{\sigma}_{i}^{z}\hat{\sigma}_{j}^{z}
  - h_{\text{eff}} \sum_{i=1}^{N}\hat{\sigma}_{i}^{z} \right]
  - \sqrt{g_{\text{u}}g_{\text{l}}}\frac{1-s}{c}\sum_{i=1}^{N}\hat{\sigma}_{i}^{x}.
  \label{WP_eq:apdx_H_equiv_spin-1/2}
\end{align}
Here, the effective longitudinal field is given as
\begin{align}
  h_{\text{eff}} = h + \frac{1-s}{s}\frac{g_{\text{u}}-g_{\text{l}}}{2c}\omega.
  \label{WP_eq:apdx_quantum_heff}
\end{align}

\bibliography{Reference}

\begin{thebibliography}{10}

\bibitem{kadowaki1998quantumannealing}
T.~Kadowaki and H.~Nishimori: Phys. Rev. E {\bfseries 58} (1998) 5355.

\bibitem{farhi2000quantum}
E.~Farhi, J.~Goldstone, S.~Gutmann, and M.~Sipser: arXiv preprint
  quant-ph/0001106  (2000).

\bibitem{farhi2001aquantum}
E.~Farhi, J.~Goldstone, S.~Gutmann, J.~Lapan, A.~Lundgren, and D.~Preda:
  Science {\bfseries 292} (2001) 472.

\bibitem{morita2008mathematical}
S.~Morita and H.~Nishimori: Journal of Mathematical Physics {\bfseries 49}
  (2008) 125210.

\bibitem{santoro2006optimization}
G.~E. Santoro and E.~Tosatti: Journal of Physics A: Mathematical and General
  {\bfseries 39} (2006) R393.

\bibitem{katzgraber2015seeking}
H.~G. Katzgraber, F.~Hamze, Z.~Zhu, A.~J. Ochoa, and H.~Munoz-Bauza: Phys. Rev.
  X {\bfseries 5} (2015) 031026.

\bibitem{denchev2016what}
V.~S. Denchev, S.~Boixo, S.~V. Isakov, N.~Ding, R.~Babbush, V.~Smelyanskiy,
  J.~Martinis, and H.~Neven: Phys. Rev. X {\bfseries 6} (2016) 031015.

\bibitem{das2008colloquim}
A.~Das and B.~K. Chakrabarti: Rev. Mod. Phys. {\bfseries 80} (2008) 1061.

\bibitem{tanaka2017quantum}
S.~Tanaka, R.~Tamura, and B.~K. Chakrabarti: {\em Quantum spin glasses,
  annealing and computation} (Cambridge University Press, 2017).

\bibitem{johnson2011quantum}
M.~W. Johnson, M.~H. Amin, S.~Gildert, T.~Lanting, F.~Hamze, N.~Dickson,
  R.~Harris, A.~J. Berkley, J.~Johansson, P.~Bunyk, et~al.: Nature {\bfseries
  473} (2011) 194.

\bibitem{maezawa2017design}
M.~Maezawa, K.~Imafuku, M.~Hidaka, H.~Koike, and S.~Kawabata: 2017 16th
  International Superconductive Electronics Conference (ISEC), June 2017, pp.
  1--3.

\bibitem{chen2011experimental}
H.~Chen, X.~Kong, B.~Chong, G.~Qin, X.~Zhou, X.~Peng, and J.~Du: Phys. Rev. A
  {\bfseries 83} (2011) 032314.

\bibitem{morita2007convergence}
S.~Morita and H.~Nishimori: Journal of the Physical Society of Japan {\bfseries
  76} (2007) 064002.

\bibitem{jansen2007bounds}
S.~Jansen, M.-B. Ruskai, and R.~Seiler: Journal of Mathematical Physics
  {\bfseries 48} (2007) 102111.

\bibitem{sachdev2011quantum}
S.~Sachdev: {\em Quantum phase transitions} (Cambridge university press, 2011).

\bibitem{znidaric2006exponential}
M.~\ifmmode \check{Z}\else \v{Z}\fi{}nidari\ifmmode~\check{c}\else \v{c}\fi{}
  and M.~Horvat: Phys. Rev. A {\bfseries 73} (2006) 022329.

\bibitem{jorg2010energy}
T.~J{\"o}rg, F.~Krzakala, J.~Kurchan, A.~C. Maggs, and J.~Pujos: EPL
  (Europhysics Letters) {\bfseries 89} (2010) 40004.

\bibitem{jorg2010firstorder}
T.~J{\"o}rg, F.~Krzakala, G.~Semerjian, and F.~Zamponi: Phys. Rev. Lett.
  {\bfseries 104} (2010) 207206.

\bibitem{dusuel2005continuous}
S.~Dusuel and J.~Vidal: Phys. Rev. B {\bfseries 71} (2005) 224420.

\bibitem{seki2012quantum}
Y.~Seki and H.~Nishimori: Physical Review E {\bfseries 85} (2012) 051112.

\bibitem{seoane2012many}
B.~Seoane and H.~Nishimori: Journal of Physics A: Mathematical and Theoretical
  {\bfseries 45} (2012) 435301.

\bibitem{seki2015quantum}
Y.~Seki and H.~Nishimori: Journal of Physics A: Mathematical and Theoretical
  {\bfseries 48} (2015) 335301.

\bibitem{susa2018exponential}
Y.~Susa, Y.~Yamashiro, M.~Yamamoto, and H.~Nishimori: Journal of the Physical
  Society of Japan {\bfseries 87} (2018) 023002.

\bibitem{wajnflasz1971transitions}
J.~Wajnflasz and R.~Pick: Journal de Physique Colloques {\bfseries 32} (1971)
  C1.

\bibitem{bousseksou1992ising}
A.~Bousseksou, J.~Nasser, J.~Linares, K.~Boukheddaden, and F.~Varret: Journal
  de Physique I {\bfseries 2} (1992) 1381.

\bibitem{bousseksou1993ising}
A.~Bousseksou, F.~Varret, and J.~Nasser: Journal de Physique I {\bfseries 3}
  (1993) 1463.

\bibitem{miyashita2010phase}
S.~Miyashita: Proceedings of the Japan Academy, Series B {\bfseries 86} (2010)
  643.

\bibitem{miyashita2003generalized}
S.~Miyashita and N.~Kojima: Progress of Theoretical Physics {\bfseries 109}
  (2003) 729.

\bibitem{miyashita2005structures}
S.~Miyashita, Y.~Konishi, H.~Tokoro, M.~Nishino, K.~Boukheddaden, and
  F.~Varret: Progress of Theoretical Physics {\bfseries 114} (2005) 719.

\bibitem{tokoro2006huge}
H.~Tokoro, S.~Miyashita, K.~Hashimoto, and S.-i. Ohkoshi: Phys. Rev. B
  {\bfseries 73} (2006) 172415.

\bibitem{nishimori2010elements}
H.~Nishimori and G.~Ortiz: {\em Elements of phase transitions and critical
  phenomena} (OUP Oxford, 2010).

\bibitem{suzuki1976relationship}
M.~Suzuki: Progress of Theoretical Physics {\bfseries 56} (1976) 1454.

\bibitem{oliphant2006guide}
T.~E. Oliphant: {\em A guide to NumPy} (Trelgol Publishing USA, 2006), Vol.~1.

\bibitem{scipy}
E.~Jones, T.~Oliphant, P.~Peterson, et~al.
\newblock {SciPy}: Open source scientific tools for {Python}, 2001--.
\newblock [Online; accessed 2018-12-14].

\bibitem{hunter2007}
J.~D. Hunter: Computing In Science \& Engineering {\bfseries 9} (2007) 90.

\end{thebibliography}

\end{document}